\documentclass[a4paper,11pt,preprintnumbers]{article}
\pdfoutput=1

\usepackage{amssymb,mathrsfs,amsmath}
\usepackage{a4wide}
\usepackage{color,xcolor}
\usepackage{slashed,soul}
\usepackage{graphicx}
\usepackage{amsfonts}
\usepackage{cancel}
\usepackage{lscape}
\def\linkcolor{cyan!70!black}

\usepackage[
colorlinks=true
,urlcolor=\linkcolor
,anchorcolor=\linkcolor
,citecolor=\linkcolor
,filecolor=\linkcolor
,linkcolor=\linkcolor
,menucolor=\linkcolor
,linktocpage=true
,pdfproducer=medialab
,pdfa=true
]{hyperref}
\usepackage{amsthm}
\usepackage{booktabs}
\usepackage{array}
\usepackage{rotating}
\usepackage[numbers, sort&compress]{natbib}
\usepackage{multirow}
\usepackage{float}
\usepackage[utf8]{inputenc}
\usepackage[T1]{fontenc}
\usepackage{appendix}
\usepackage{epsfig}
\usepackage{orcidlink}
\usepackage{comment}
\usepackage{tabularx}

\newcommand{\beq}{\begin{equation}} 
\newcommand{\eeq}{\end{equation}} 
\newcommand{\ba}{\begin{array}}  
\newcommand{\ea}{\end{array}} 
\newcommand{\bea}{\begin{eqnarray}}  
\newcommand{\eea}{\end{eqnarray} }  
\newcommand{\bal}{\begin{align}}
\newcommand{\eal}{\end{align}}   
\newcommand{\bi}{\begin{itemize}}  
\newcommand{\ei}{\end{itemize}}  
\newcommand{\ben}{\begin{enumerate}}  
\newcommand{\een}{\end{enumerate}}  
\newcommand{\bc}{\begin{center}}
\newcommand{\ec}{\end{center}} 
\newcommand{\bt}{\begin{table}}
\newcommand{\et}{\end{table}}  
\newcommand{\btb}{\begin{tabular}}
\newcommand{\etb}{\end{tabular}}

\def\arrvline{\hfil\kern\arraycolsep\vline\kern-\arraycolsep\hfilneg}

\newcolumntype{Y}{>{\centering\arraybackslash}X}

\allowdisplaybreaks

\let\OLDthebibliography\thebibliography
\renewcommand\thebibliography[1]{
  \OLDthebibliography{#1}
  \setlength{\parskip}{0pt}
  \setlength{\itemsep}{0pt plus 0.3ex}
}

\begin{document}

\vspace{1cm}

\begin{titlepage}

\begin{flushright}
IFT-UAM/CSIC-25-20\\
FTUV-25-0225.3158
 \end{flushright}
\vspace{0.2truecm}

\begin{center}
\renewcommand{\baselinestretch}{1.8}\normalsize
\boldmath
{\LARGE\textbf{
Misconceptions in Neutrino Oscillations \\ in presence of non-Unitary Mixing}}
\unboldmath
\end{center}

\vspace{0.4truecm}

\renewcommand*{\thefootnote}{\fnsymbol{footnote}}

\begin{center}

{
Mattias Blennow$^1$\footnote{\href{mailto:emb@kth.se}{emb@kth.se}}\orcidlink{0000-0001-5948-9152},
Pilar Coloma$^2$\footnote{\href{mailto:pilar.coloma@ift.csic.es}{pilar.coloma@ift.csic.es}}\orcidlink{0000-0002-1164-9900}, 
Enrique Fern\'andez-Mart\'inez$^2$\footnote{\href{mailto:enrique.fernandez@csic.es}{enrique.fernandez@csic.es}}\orcidlink{0000-0002-6274-4473}, \\[0.5ex]
Josu Hern\'andez-Garc\'ia$^3$\footnote{\href{mailto:josu.hernandez@ific.uv.es}{josu.hernandez@ific.uv.es}}\orcidlink{0000-0003-0734-0879},
Jacobo L\'opez-Pav\'on$^3$\footnote{\href{mailto:Jacobo.Lopez@uv.es}{jacobo.lopez@uv.es}}\orcidlink{0000-0002-9554-5075},
Xabier Marcano$^2$\footnote{\href{mailto:xabier.marcano@uam.es}{xabier.marcano@uam.es}}\orcidlink{0000-0003-0033-0504},\\[0.5ex]
Daniel Naredo-Tuero$^2$\footnote{\href{mailto:daniel.naredo@ift.csic.es}{daniel.naredo@ift.csic.es}}\orcidlink{0000-0002-5161-5895}
and Salvador Urrea$^{4}$\footnote{\href{mailto:salvador.urrea@ijclab.in2p3.fr}{salvador.urrea@ijclab.in2p3.fr}}\orcidlink{0000-0002-7670-232X}
}

\vspace{0.7truecm}

{\footnotesize
$^1$Department of Physics, School of Engineering Sciences, KTH Royal Institute of Technology, \\
AlbaNova University Center, Roslagstullsbacken 21, SE-106 91 Stockholm, Sweden
\\[.5ex]
$^2$Instituto de F\'{\i}sica Te\'orica UAM/CSIC,\\
Universidad Aut\'onoma de Madrid, Cantoblanco, 28049 Madrid, Spain
\\[.5ex]
$^3$ Instituto de F\'{\i}sica Corpuscular, Universidad de Valencia and CSIC\\ 
 Edificio Institutos Investigaci\'on, Catedr\'atico Jos\'e Beltr\'an 2, 46980 Spain\\
$^4$ IJCLab, Pôle Théorie (Bat. 210), CNRS/IN2P3, 91405 Orsay, France
}

\vspace*{2mm}
%\today
\end{center}

\renewcommand*{\thefootnote}{\arabic{footnote}}
\setcounter{footnote}{0}

%\vspace{0.3cm}
\begin{abstract}
Deviations from unitarity of the CKM matrix in the quark sector are considered excellent windows to probe physics beyond the Standard Model. In its leptonic counterpart, the PMNS matrix, these searches are particularly motivated, as the new physics needed to generate neutrino masses often leads to non-unitary mixing among the standard neutrinos. It is then interesting to consider how neutrino oscillations are affected in such scenario. 
This simple question is, however, subject to several subtleties: What is the correct way to define oscillation probabilities for a non-unitary mixing matrix? Do these probabilities add up to one? Does a non-unitary mixing matrix lead to observable flavor transitions at zero distance? What is the interplay between unitarity constraints obtained from neutrino oscillations and from electroweak precision data? This work aims to shed light on these issues and to clarify the corresponding misconceptions commonly found in the literature. We also compile updated bounds from neutrino oscillation searches to compare with those from flavour and electroweak precision observables. 
 
\end{abstract}

\end{titlepage}

\tableofcontents

%%%%%%%%%%%%%%%%%%%%%%%%%%%
\section{Introduction}

The phenomenon of flavour change in the Standard Model (SM) of particle physics is entirely controlled by the unitary $3\times 3$ CKM mixing matrix. These transitions are suppressed by the celebrated GIM mechanism~\cite{Glashow:1970gm} and are therefore excellent probes of new physics, since this suppression would, in general, not apply. As such, unitarity tests of the CKM matrix are a fundamental tool to search for physics beyond the SM. 

The discovery of neutrino masses and mixings through the observation of neutrino oscillations implies the existence of the PMNS matrix, the leptonic analogue of the CKM. Global fits to the available neutrino oscillation data already constrain with very good accuracy its mixing angles and start to provide constraints on its CP-violating phase~\cite{Esteban:2024eli,Capozzi:2021fjo,deSalas:2020pgw}. It is then natural to consider how this picture would be affected in presence of a non-unitary (NU) mixing matrix. Indeed, this search is particularly well-motivated in the neutrino sector since neutrino masses already require physics beyond the SM. Moreover, the simplest SM extension to explain their origin, {\it i.e.}~the inclusion of right-handed neutrinos as in the popular type-I Seesaw mechanism~\cite{Minkowski:1977sc,Mohapatra:1979ia,Yanagida:1979as,Gell-Mann:1979vob}, already predicts unitarity deviations of the PMNS mixing matrix at some level~\cite{Broncano:2002rw}. Remarkably, even if only neutrino oscillation experiments are presently sensitive to the individual PMNS matrix elements, weak interaction rates would be affected in presence of non-unitarity and thus unitarity deviations can also be very efficiently probed via flavour and electroweak precision observables (EWPO)~\cite{Petcov:1976ff,Bilenky:1977du,Cheng:1977vn,Marciano:1977wx,Lee:1977qz,Lee:1977tib,Shrock:1980vy,Schechter:1980gr,Shrock:1980ct,Shrock:1981wq,Langacker:1988ur,Pilaftsis:1992st,Ilakovac:1994kj,Nardi:1994iv,Tommasini:1995ii,Illana:2000ic,Loinaz:2003gc,Arganda:2004bz,Loinaz:2004qc,Antusch:2006vwa,Antusch:2008tz,Biggio:2008in,Alonso:2012ji,Abada:2012mc,Akhmedov:2013hec,Basso:2013jka,Abada:2013aba,Arganda:2014dta,Antusch:2014woa,Antusch:2015mia,Abada:2014cca,Abada:2015oba,Abada:2015trh,Fernandez-Martinez:2015hxa,Fernandez-Martinez:2016lgt,Abada:2016awd,DeRomeri:2016gum,Arganda:2016zvc,Herrero:2018luu,Marcano:2019rmk,Chrzaszcz:2019inj,Coutinho:2019aiy,Calderon:2022alb,Blennow:2023mqx}. 

There is by now a very large body of works in the literature dealing with the impact of a non-unitary PMNS mixing matrix at present or future neutrino oscillation facilities~\cite{Bilenky:1992wv,Bergmann:1998rg,Czakon:2001em,Bekman:2002zk,Antusch:2006vwa,Fernandez-Martinez:2007iaa,Goswami:2008mi,Antusch:2009pm,Meloni:2009cg,Xing:2011ur,Forero:2011pc,Qian:2013ora,Parke:2015goa,Escrihuela:2015wra,Blennow:2016jkn,Ge:2016xya,Escrihuela:2016ube,Miranda:2016wdr,Pas:2016qbg,Dutta:2016czj,Bielas:2017lok,Miranda:2018yym,Martinez-Soler:2018lcy,Li:2018jgd,Soumya:2018nkw,DeGouvea:2019kea,Escrihuela:2019mot,Miranda:2019ynh,Ellis:2020ehi,Ellis:2020hus,Miranda:2020syh,Chakraborty:2020brc,Hu:2020oba,Coloma:2021uhq,Forero:2021azc,Denton:2021mso,Denton:2021rsa,Agarwalla:2021owd,Denton:2022pxt,Fong:2022oim,Aloni:2022ebm,Fong:2023fpt,Fong:2023ams,Celestino-Ramirez:2023zox,Sahoo:2023mpj,Kozynets:2024xgt,Trzeciak:2025hap}. This issue is, however, riddled with many subtleties that have led to a number of misconceptions which have propagated through the literature. Indeed, as we will elaborate later, the very definition of oscillation probability is not completely straightforward when in presence of a non-unitary mixing matrix and the comparison with what is experimentally determined is also subtle. In this work we aim to shed light and clarify these misconceptions and potentially incorrect results. In particular, we will first cover the possible definitions of oscillation probability and their properties. We will then address a number of subtle issues that have led to misconceptions in previous works, in particular:
\begin{itemize}

\item We comment on the apparent paradox of oscillation probabilities not adding up to one when in presence of non-unitary mixing.

\item We discuss how oscillation probabilities generally need to be ``normalized'' so as to match what is experimentally determined and how this in turn impacts the observability of some non-unitarity effects that might have been naively expected. Notably ``zero-distance effects'' in disappearance channels.

\item We comment on the option of employing unitarity triangles, as common practice in the quark sector, to study non-unitary in the lepton sector. We find that this tool is not optimal given the differences between the two sectors, which we discuss in detail.

\item We remark on the impact of the usual practice of extracting the mixing parameters as input for the non-unitary mixing matrix from analyses where its unitarity is assumed. 
This procedure is at the root of seemingly inconsistent results when analyzing non-unitarity effects through different parametrizations.  

\item We discuss how the usual extraction of $G_F$ from $\mu$ decay is affected in presence of a non-unitary mixing matrix and how this indirect dependence should also be taken into account when studying matter effects in neutrino propagation.

\item We discuss what is the correct description of the number of events detected via neutral current interactions in the context of a non-unitarity mixing matrix, which is particularly subtle.

\item  We present updated constraints on non-unitarity stemming from neutrino oscillation data and compare them to those from flavour and electroweak precision observables, discussing their relative importance and possible complementarity.

\end{itemize}

Finally, we also discuss how the presence of sterile neutrinos and in a regime where they are light enough to be produced at the neutrino experiment but too heavy for their oscillations to be resolved by the detector, displays very similar phenomenology to that of a non-unitary mixing matrix. However, in this regime, the bounds from flavour and electroweak precision observables do not apply, making the oscillation probes the main experimental avenue to search for their signals.

\section{What is non-unitarity and where is it coming from?}
\label{sec:nouni}

The scenario we will consider and refer to as non-unitarity\footnote{Some works, such as~\cite{Parke:2015goa}, consider instead a different regime with light sterile neutrinos but whose oscillations are averaged out. We will discuss this regime and its connection with the alternative definition at the end of Section~\ref{sec:bounds}. For more details see Refs.~\cite{Blennow:2016jkn,Fong:2016yyh}.} is characterized by a relation between the neutrino fields $\nu_i$ that diagonalize their mass matrix and the neutrino flavour fields $\nu_\alpha$, in which the weak interactions are diagonal, given by a non-unitary leptonic mixing $N$ such that:
\begin{equation}
\nu_\alpha = N_{\alpha i}\, \nu_i\,,
\label{eq:N}
\end{equation}
with $\alpha = e, \mu, \tau$ and $i=1,2,3$, and a sum over repeated indices is assumed. This implies that neutrino CC and NC interactions are modified and given by:
\begin{equation}
    \mathcal{L} \supset-\frac{g}{\sqrt{2}}\left(W_\mu^{-} \bar{\ell}_{\alpha} \gamma_\mu P_L N_{\alpha i} \nu_{i}+\text { h.c.}\right)-\frac{g}{2 \cos \theta_W}Z_\mu \bar{\nu}_{i} \gamma^\mu P_L\left(N^{\dagger} N\right)_{i j} \nu_{j}\,.
    \label{eq:lagr}
\end{equation}

The simplest SM extension leading to this scenario corresponds to introducing new sterile (right-handed) neutrinos mixing with the active SM ones, such that the $N$ matrix is just a sub-block of a larger unitary matrix $\mathcal U$. In more general terms, this non-unitarity framework is equivalent to introducing only the following dimension-6 operator:
\begin{equation}
    \delta\mathcal{L}^{\rm d=6}=\frac{4 \eta_{\alpha \beta}}{v^2} \left(\overline{L}_{\alpha} \tilde{H}\right)i\vec{\cancel{\partial}
}\left(\tilde{H}^\dagger L_\beta\right),
    \label{eq:d6op}
\end{equation}
with $\eta$ an Hermitian matrix. This is in fact the sole $d=6$ operator obtained at tree level upon integrating out the heavy right-handed neutrinos~\cite{Broncano:2002rw} in a type-I Seesaw mechanism\footnote{We note that, in the heavy neutrino extension of the SM, the coefficient of this operator, $\eta$, is positive-definite while, in all generality, it is only constrained to be Hermitian.}. 
After the Higgs develops its vacuum expectation value $v$, the impact of this operator is to lead to non-canonical neutrino kinetic terms which, upon diagonalization and normalization, leads to the desired non-unitary leptonic mixing matrix as~\cite{Broncano:2002rw}:
\begin{equation}
N = (I-\eta)V\,,
\label{eq:eta_parametrisation}
\end{equation}
where $V$ is a unitary matrix. This parametrization is completely general~\cite{Fernandez-Martinez:2007iaa}, as any matrix can be decomposed as the product of a Hermitian and a unitary matrix, and it has the advantage of providing a natural connection with the relevant effective operator in Eq.~\eqref{eq:d6op}.

Another popular parametrization, and particularly convenient for the study of neutrino oscillations, is through the product of a lower triangular matrix $\alpha$ times a (different) unitary matrix $U$~\cite{Xing:2007zj,Escrihuela:2015wra}:
\begin{equation}
    N=\left(I-\alpha\right)U\,,
    \label{eq:alpha_parametrisation}
\end{equation}
where
\begin{equation}
    \alpha=
    \begin{pmatrix}
        \alpha_{ee} & 0 & 0\\
        \alpha_{\mu e} & \alpha_{\mu\mu} & 0\\
        \alpha_{\tau e}&\alpha_{\tau\mu} & \alpha_{\tau\tau}
    \end{pmatrix}.
\end{equation}
At leading order in the small parameters that encode the unitarity deviations ($\eta$ or $\alpha$) the relation between the two parametrizations is given by~\cite{Blennow:2016jkn}:
\begin{equation}
\alpha_{\alpha\alpha}=\eta_{\alpha\alpha}+\mathcal O(\eta^2)\,,\hspace{0.5cm}\alpha_{\alpha\beta}=2\eta_{\alpha\beta}+\mathcal O(\eta^2)\,,\,\,\,(\alpha\neq\beta)\,.
\end{equation}
The relation between $U$ and $V$ can also be found in Ref.~\cite{Blennow:2016jkn}. Notice that neither $U$ nor $V$ correspond exactly to the ``standard'' unitary PMNS matrix through which neutrino oscillations are normally analized and whose mixing angles are obtained from global fits to the 3-neutrino paradigm~\cite{Esteban:2024eli,Capozzi:2021fjo,deSalas:2020pgw}. 
We will elaborate on this issue in Section~\ref{sec:parametrizations}. We also discuss options that do not rely on a matrix that could be identified with the unitary PMNS~\cite{Parke:2015goa,Ellis:2020ehi,Ellis:2020hus} in Section~\ref{sec:triangles}.

\section{Defining Probabilities}
\label{sec:probabilities}

The relation in Eq.~\eqref{eq:N} implies that the flavour basis in which neutrinos are produced and detected is not orthonormal. Conversely, the mass basis is necessarily orthonormal since it diagonalizes the Hermitian matrix $M^\dagger M$, with $M$ the mass matrix.
As such, several subtle issues arise that are not intuitive and often lead to misconceptions. First and foremost among these is the key fact that oscillation probabilities are not directly observable but rather indirectly inferred from numbers of events and therefore different definitions are possible and have been used in the literature. In particular, we can define a ``theoretical probability'' $P$ in the usual manner, as the squared amplitude for the transition between two quantum states. With this definition, this ``theoretical probability'' is a true probability but, in general, might not directly map to the measured quantity. Conversely, we can define an ``experimental probability'' $\tilde{P}$ as the one extracted by a given experiment from the comparison of a measured number of events to its expected value in the absence of oscillations. This, by definition, is observable but, in general, may not correspond to an actual probability. In the following, we discuss both definitions and their relation in more detail, highlighting when the two coincide and when they differ.  

\subsection{Theoretical probability}
\label{sec:theoretical_probabilities}

In order to obtain properly normalized transition probabilities, the production and detection states from CC interactions ({\it i.e.}~the flavour eigenvalues $| \nu_\alpha \rangle $) need to be normalized such that:
\begin{equation}
\left| \nu_\alpha \right> = \frac{ N_{\alpha i}^* \left| \nu_i \right>}{\sqrt{\sum_i |N_{\alpha i}|^2}}\,.
\label{eq:norm}
\end{equation}
And thus the ``theoretical probability'' $P$ of obtaining a state $\left| \nu_\beta \right>$ when starting from a $\left| \nu_\alpha \right>$ after propagating for a baseline $L$ will be simply given by:
\begin{equation}
P_{\nu_\alpha \to \nu_\beta}(L) = \big|\left< \nu_{\beta}| \nu_\alpha(L) \right>\big|^2
= \dfrac{\left\lvert\sum_{i} N_{\alpha i}^*\,e^{-i m_{i}^2 L/2E} N_{\beta i}\right\rvert^2}{\sum_i\left\lvert N_{\alpha i}\right\rvert^2 \sum_j\left\lvert N_{\beta j}\right\rvert^2} \,,
\label{eq:theoprob}
\end{equation}
where propagation in vacuum has been assumed for simplicity.

As already mentioned, this definition yields true probabilities. Nevertheless, although the states in Eq.~\eqref{eq:norm} are normalized, since $N$ is non-unitarity they are not orthogonal. Thus:
\begin{equation}
\left< \nu_\beta | \nu_\alpha \right> \neq \delta_{\alpha \beta}\,,
\label{eq:zerodist}
\end{equation}
and these probabilities will generally not add up to one, since the different elements of the basis are not mutually exclusive. This fact has led to some misconceptions with alternative definitions of ``probabilities'' adding up to one being proposed in the literature. However, these alternative definitions are not correct. Actual probabilities should \emph{not} add up to one when describing events that are \emph{not} mutually exclusive. 

As a toy example, let us consider a basis for the spin states of an electron where the first element is positive spin in the X direction and the second element is positive spin in the Z direction. The two elements of the basis are not orthogonal and, therefore, do not represent mutually exclusive outcomes. But they do form a complete and normalized basis in complete analogy to our neutrino flavour basis. Now consider a state which has positive spin along the X axis. The probability to detect it as the first element of our basis is 1 and the probability to detect it as the second element of our basis is $1/2$. The probabilities add up to more than 1 but they are clearly correct and it is simply a consequence of the elements of the basis not being mutually exclusive. Similarly, if the initial state had negative spin along the X axis, the probabilities to detect it as the elements of the basis would add up to less than 1, but there is nothing inconsistent with this observation. 

Another consequence of the non-orthogonality of the flavour basis is the presence of ``zero distance effects'', where a flavour transition does not need neutrinos to propagate. We will discuss them in detail in Section~\ref{sec:zerodistance}.

\subsection{Experimental probability}
\label{sec:experimental_probabilities}

Even if the probabilities defined in the previous section do represent transition probabilities between two given states, it is important to notice that sometimes they do not correspond to what is experimentally determined. The actual measurement rather corresponds to the number of oscillated events from which collaborations extract the ``oscillation probabilities'' by comparing them with \emph{estimated} number of events in absence of oscillations. These estimations necessarily require the input of the neutrino flux and cross section that will in turn be based on some observable that will also be affected by the non-unitarity of the mixing matrix. 

Indeed, since neutrinos will always be produced or detected via the weak interactions of Eq.~\eqref{eq:lagr}, the decay rates in which neutrinos are produced or the cross sections where they are detected associated to a charged lepton $\ell_\alpha$ will be proportional to $\sum_i |N_{\alpha i}|^2 \neq 1$, and thus depend on the non-unitarity parameters. This indirect dependence must be propagated and considered in the oscillation probabilities measured in this way, which reduces the sensitivity to non-unitarity effects in many experiments~\cite{Blennow:2016jkn,Aloni:2022ebm,Coloma:2024ict}. Indeed, denoting the event rates measured in presence of oscillations by $R_{\alpha\to\beta}$, the ``experimental probability'' can be extracted as: 
\begin{equation}
\tilde{P}_{\nu_\alpha \to \nu_\beta}(L) = \dfrac{\frac{d R_{\alpha\to\beta}}{dE}}{\mathcal{N}_{\text{tgt}}\varepsilon\frac{d \phi_\alpha}{d E} \sigma_\beta}
= \dfrac{\left\lvert\sum_{i} N_{\alpha i}^*\,e^{-i m_{i}^2 L/2E} N_{\beta i}\right\rvert^2}{\sum_i\left\lvert N_{\alpha i}\right\rvert^2 \sum_j\left\lvert N_{\beta j}\right\rvert^2},
\label{eq:noneardet}
\end{equation}
where $\phi_\alpha$ and $\sigma_\beta$ are the flux and cross section used to estimate that rate, $\mathcal{N}_{\text{tgt}}$ is the number of targets and $\varepsilon$ the detection efficiency. Note that in practice, this is often tantamount to normalizing the neutrino states as in Eq.~\eqref{eq:norm}, so it often coincides with the theoretical probability in Eq.~\eqref{eq:theoprob}. The exception is when, for some reason, the estimation is performed through a process involving a different flavour than the one considered in the transition. This can for example be the case of near detectors for oscillation facilities looking for the $\nu_\mu \to \nu_e$ oscillation when the flux and cross section are estimated from measurements of $\nu_\mu$ at the near detector. In this example the ``experimental probability'' $\tilde P$ extracted by correlating the number of events between the far and near detectors would rather correspond to~\cite{Blennow:2016jkn}: 
\begin{equation}
\tilde{P}_{\nu_\mu \to \nu_e}(L)= \dfrac{\left\lvert\sum_{i} N_{\mu i}^*\,e^{-i m_{i}^2 L/2E} N_{e i}\right\rvert^2}{\sum_i\left\lvert N_{\mu i}\right\rvert^2 \sum_j\left\lvert N_{\mu j}\right\rvert^2} = \dfrac{\sum_i\left\lvert N_{e i}\right\rvert^2}{\sum_i\left\lvert N_{\mu i}\right\rvert^2}P_{\nu_\mu \to \nu_e}(L)\,.
\label{eq:neardet}
\end{equation}
Thus, in this particular example, the ``theoretical'' and the ``experimental'' probabilities do not coincide. 

It is important to notice that, in practice, there will \textit{always} be two normalization factors as in the middle expression of Eqs.~\eqref{eq:noneardet} and~\eqref{eq:neardet}. One corresponding to propagating the dependence on~$N_{\alpha i}$ in the production process and another for the detection interaction. It has been argued on occasions that such normalization factor would not appear if the estimation of the neutrino flux or cross section used to extract the ``experimental'' oscillation probability does not utilize any actual measurement but rather relies only in ``pure'' SM predictions or processes not involving neutrinos (and hence not affected by~$N_{\alpha i}$). While this is technically correct, in practice it \emph{never happens}. Even when the meson flux produced in a beam experiment or atmospheric shower is estimated via hadroproduction measurements and MonteCarlos, the neutrino flux will be obtained from that meson flux times the corresponding branching ratio to~$\nu_i$. In order for these branching ratios to add up to one, the normalization factor $\sum_i |N_{\alpha i}|^2$ is needed~\cite{Aloni:2022ebm} in the denominator. In presence of non-unitarity, there can be no truly ``pure'' SM prediction for a neutrino flux or cross section via charged currents. Indeed, this would require, at the very least and avoiding more direct measurements, the input of the CKM matrix element $V_{ud}$, which is determined via $\beta$ or meson decays involving neutrinos. Hence, even in such a scenario a $\sum_i |N_{\alpha i}|^2$ factor would appear.

An observation that can be useful to understand why 
the two normalization factors in Eqs.~\eqref{eq:noneardet} and~\eqref{eq:neardet} are always necessary is that, in a toy scenario where the effect of non-unitarity was simply an overall normalization $n$ such that $N_{\alpha i} = n U_{\alpha i}$ with $U$ unitary, it would be impossible to distinguish this scenario from the SM rescaling the coupling constant $g \to n\,g $ as long as only $\nu$ CC interactions are considered\footnote{Unless processes with different number of neutrino legs are compared. For instance comparing measurements of $G_F$ via $\mu$ decay (2 neutrino legs), via $\beta$ decays (1 neutrino leg) or via $M_W$ (no neutrino legs). This provides very good sensitivity to non-unitarity, as discussed in Section~\ref{sec:bounds}, but it is unrelated to neutrino oscillations.}, which is mostly the case for neutrino oscillations. Thus, the same number of powers of $N$ needs to be present in the numerator and denominator of Eqs.~\eqref{eq:noneardet} and~\eqref{eq:neardet}. Otherwise, spurious sensitivity to $n$ would have been introduced via an incorrect normalization. This oversight is rather common in the existing literature.

\section{Common misconceptions}

\subsection{The zero-distance effect}
\label{sec:zerodistance}

The necessary normalization of the oscillation probability has the important consequence of canceling some of the dependence on unitarity deviations that could have been naively expected. Indeed, when the estimation of the unoscillated events is done using data with the same flavour as the targeted transition, the ``experimental'' probability in vacuum is given by:
%Indeed, when the estimation of the unoscillated events is done from data with the same flavours than the targeted transition, the ``experimental'' probability in vacuum is given by:
\begin{align}
    \tilde{P}_{\nu_\alpha\to\nu_\beta}(L) 
    = \frac{1}{1-2\alpha_{\alpha\alpha}-2\alpha_{\beta\beta}}&
    \bigg\lvert (1-\alpha_{\alpha\alpha} - \alpha_{\beta\beta}) \sum\limits_{i}U_{\alpha i}^*\,e^{-i \frac{m_{i}^2 L}{2E}} U_{\beta i}  \nonumber \\ 
    & - \sum\limits_{i,\gamma\neq\alpha}\alpha_{\alpha\gamma}^* U_{\gamma i}^* e^{-i \frac{m_{i}^2 L}{2E}} U_{\beta i}-\sum\limits_{i,\gamma\neq\beta} U_{\alpha i}^* e^{-i \frac{m_{i}^2 L}{2E}} \alpha_{\beta\gamma}U_{\gamma i} \bigg\rvert^2 
    + \mathcal{O}(\alpha^2) \nonumber \\
     = \bigg\lvert \sum_{i} U_{\alpha i}^*e^{-i \frac{m_{i}^2 L}{2E}}U_{\beta i}  &-\sum_{i,\gamma\neq\alpha}\alpha_{\alpha\gamma}^* U_{\gamma i}^* e^{-i \frac{m_{i}^2 L}{2E}} U_{\beta i}-\sum_{i,\gamma\neq\beta} U_{\alpha i}^* e^{-i \frac{m_{i}^2 L}{2E}} \alpha_{\beta\gamma}U_{\gamma i} \bigg\rvert^2 + \mathcal O(\alpha^2)\,,
\label{eq:normalised_probability}
\end{align}
and the leading order dependence on the flavour-diagonal non-unitary parameters cancels altogether. This effect is especially relevant for zero-distance effects in the disappearance channel, where,  without the proper normalization, one would obtain a linear dependence on the flavour-diagonal non-unitary parameters for the disappearance channels. As such, these data have been often considered a source of strong constraints on the diagonal entries encoding unitarity deviations. However, particularizing Eq.~\eqref{eq:noneardet} for $\alpha=\beta$ and $L=0$, the sensitivity to $\alpha$ is lost at all orders: 
\begin{equation}
\tilde{P}_{\nu_\alpha\to\nu_\alpha}(L=0)= \frac{\left\lvert\sum_{i} N_{\alpha i}^* N_{\alpha i}\right\rvert^2}{\sum_i\left\lvert N_{\alpha i}\right\rvert^2 \sum_j\left\lvert N_{\alpha j}\right\rvert^2} =1 \,.
\label{eq:verytrivial}
\end{equation}
Conversely, this cancellation does not take place for $\alpha\neq\beta$,
\begin{equation}
\tilde{P}_{\nu_\alpha\to\nu_\beta}(L=0)=|\alpha_{\alpha \beta}|^2+ \mathcal O(\alpha^3)\,,
\end{equation}
and therefore the zero distance effect in appearance channels can be a very useful tool to probe for non-unitarity.
In fact, these zero-distance effects have been stringently constrained by, for example, KARMEN~\cite{Eitel:2000by} and NOMAD~\cite{NOMAD:2001xxt,NOMAD:2003mqg}.

\subsection{Non-unitarity parametrizations and comparison with oscillation data}
\label{sec:parametrizations}

In Ref.~\cite{Blennow:2016jkn} it was pointed out that the oscillation probabilities in presence of unitarity deviations appear very different when expressed using different parametrizations such as those introduced in Eqs.~\eqref{eq:eta_parametrisation}-\eqref{eq:alpha_parametrisation}. This apparent contradiction is solved through the realization that the unitary matrices involved in both of them, $U$ and $V$, are not the same. Moreover, they generally will not coincide either with what is usually dubbed the PMNS matrix, parametrized through its three mixing angles $\theta_{12}$, $\theta_{23}$ and $\theta_{13}$ and the CP violating phase $\delta$, and whose values are extracted from the results of oscillation experiments and global fits to them \textit{when assuming that the mixing matrix is unitary}~\cite{Esteban:2024eli,Capozzi:2021fjo,deSalas:2020pgw}. 

Indeed, in order to consistently constrain the non-unitarity parameters, a simultaneous fit to them as well as the mixing angles contained in the unitary part of Eqs.~\eqref{eq:eta_parametrisation}-\eqref{eq:alpha_parametrisation} should be performed. In fact, this effect is in general linear in the non-unitarity parameters (see Ref.~\cite{Blennow:2016jkn}) and, as such, it is not consistent to neglect it when studying the sensitivity to non-unitarity. This is most often overlooked in the literature, where it is not uncommon to assume the best-fit values of the standard unitary 3-flavour paradigm for the mixing angles of either $U$ or $V$, thus missing the fact that their extraction will be modified by the presence of a non-unitary matrix.\footnote{Similar conclusions apply to the light sterile neutrino scenario (see for instance~\cite{Donini:2011jh,Donini:2012tt,Goldhagen:2021kxe}).}

However, it is possible to try to optimize the non-unitarity parametrization in such a way as to minimize this non-unitarity ``contamination'' in the extraction of the mixing angles. Indeed, as previously stated, the $\alpha$-parametrization of Eq.~\eqref{eq:alpha_parametrisation} is particularly useful for this purpose. For starters, since all the entries $\alpha_{e\beta}=0$ for $\beta\neq e$, the extraction of $\theta_{13}$ from $\overline{\nu}_e$-disappearance at reactor experiments will be unchanged, since the only non-zero entry that modified production and detection ($\alpha_{ee}$) cancels altogether\footnote{Notice that the cancellation of the $\alpha$ dependence is exact at all orders in this channel even though Eq.~\eqref{eq:normalised_probability} was expanded.} due to the normalization of the probability. Following Eq.~\eqref{eq:noneardet}: 
\begin{align}
    \tilde{P}_{\overline{\nu}_e\to\overline{\nu}_e}^{\rm reactor}&=\left\lvert \sum_{i} U_{ei}e^{-\frac{i  m_{i}^2L}{2E}} U_{ei}^*-\sum_{i,\gamma\neq e} \alpha_{e\gamma} U_{\gamma i}e^{-\frac{i  m_{i}^2L}{2E}} U_{ei}^*-\sum_{i,\gamma\neq e} U_{ei} e^{-\frac{i m_{i}^2L}{2E}} \alpha^*_{e\gamma} U^*_{\gamma i}\right\rvert^2 \nonumber\\
    &= \left\lvert\sum_{i} U_{ei}e^{-\frac{i m_{i}^2L}{2E}} U_{ei}^*\right\rvert^2 \, .
    \label{eq:reactors}
\end{align}
Furthermore, since matter effects at these facilities are negligible, non-unitarity will not modify in any way the oscillation. As a consequence, the measured value of $\theta_{13}$ reported by the collaborations can be used at face value and directly identified with the $\theta_{13}$ angle entering in the definition of the unitary matrix $U$ in Eq.~\eqref{eq:alpha_parametrisation}.

Regarding the extraction of $\theta_{23}$ from $\nu_\mu$-disappearance at accelerator facilities or atmospheric neutrino oscillations, both $\alpha_{\mu\mu}$ and $\alpha_{\mu e}$ enter in the modification of the probability at production and detection. While $\alpha_{\mu\mu}$ cancels due to the normalization, $\alpha_{\mu e}$ survives at leading order. Neglecting matter effects, the oscillation probability is given by:
\begin{equation}
    \tilde{P}_{\nu_\mu\to\nu_\mu}^{\rm acc.}=\left\lvert \sum_{i} U_{\mu i}^*e^{-i \frac{m^2_i L}{2E}}  U_{\mu i}-\alpha_{\mu e}^*\sum_{i}  U_{e i}^*e^{-i \frac{m^2_i L}{2E}}U_{\mu i}-\alpha_{\mu e}\sum_{i} U_{\mu i}^* e^{-i \frac{m^2_i L}{2E}} U_{e i}+\mathcal O(\alpha^2)\right\rvert^2,
\end{equation}
where the term multiplying $\alpha_{\mu e}$, once the unitarity condition $\sum_i U_{ei}^{} U_{\mu i}^*=0$ is employed, can be rewritten as:
\begin{equation}
    \sum_i U_{ei}^*e^{-i \tfrac{m^2_{i}L}{2E}}U_{\mu i}=-e^{-i \tfrac{m^2_{1}L}{2E}}U_{e2}^{*}U_{\mu 2}\left(1-e^{-i \tfrac{\Delta m^2_{21}L}{2E}}\right)-e^{-i \tfrac{m^2_{1}L}{2E}}U_{e3}^{*}U_{\mu3}\left(1-e^{-i \tfrac{\Delta m^2_{31}L}{2E}}\right),
    \label{eq:numudis}
\end{equation}
and similarly for the term multiplying $\alpha_{\mu e}^*$. It is easy to see that the first term in the r.h.s.~of Eq.~\eqref{eq:numudis} is suppressed by $\Delta m_{21}^2L/2E$, while the second is suppressed by $\theta_{13}$. Thus, within the $\alpha$-parametrization and for vacuum oscillations, the effect of non-unitarity in production and detection in $\nu_\mu$-disappearance are subleading. Nevertheless, the non-unitarity parameters will modify the probability through the Earth matter effect, as we will discuss in Section~\ref{subsec:matter}, so this approximation is only good when matter effects are negligible. Thus, when adopting the $\alpha$-parametrization, the impact on the determination of $\theta_{23}$ is suppressed by either $\theta_{13}$, $\Delta m^2_{21}$ or the relative strength of matter effects.

Conversely, for the extraction of $\theta_{12}$ from solar neutrino experiments avoiding relevant corrections from non-unitarity is challenging and not accomplished by the $\alpha$-parametrization. Indeed, due to the sizeable matter effects in the Sun, the picture discussed above is very different since virtually all of the non-unitarity parameters modify the oscillation probability, as we will outline in Section~\ref{subsec:matter}. Consequently, the measurement of $\theta_{12}$ will be modified in a highly non-trivial manner and would require special treatment. This fact is most often ignored in the literature. On the other hand, the extraction of $\theta_{12}$ at KamLAND (and eventually JUNO) is very clean, since the same arguments as for Eq.~\eqref{eq:reactors} apply. Nevertheless, this determination is not competitive and leads to much larger uncertainties than the solar neutrino measurement for the time being. 

Having outlined some of the caveats and difficulties that are associated to the analysis of neutrino oscillations in the presence of a non-unitary mixing matrix, it should be stated that these complications may be evaded, since they are often not the best probes of the non-unitarity parameters. In fact, they may be probed in a more robust and straightforward way through the search for ``zero-distance'' flavour conversions discussed in Section~\ref{sec:zerodistance}. Alternatively, another powerful probe of these parameters is neutrino oscillations in the regime in which matter effects are dominant. This is the case for $\nu_\mu$-disappearance or even $\nu_{\tau}$-appearance for high-energy atmospheric neutrinos traversing the Earth, which is a channel best probed by neutrino observatories such as IceCube~\cite{Stuttard:2020zsj,IceCube:2024dlz} and KM3NeT~\cite{KM3NeT:2025ftj}. Since matter effects dominate over the vacuum term, the oscillation probability does not depend on the mixing angles and can be exclusively expressed in terms of the non-unitarity parameters (see e.g.~\cite{Blennow:2018hto}).

\subsection{Unitarity Triangles}
\label{sec:triangles}

Given the subtleties associated with the choice of a suitable parametrization for the unitarity deviations we just described, a potentially interesting alternative is to avoid parametrizing them altogether. Indeed, this is the standard approach when testing for CKM unitarity, where independent measurements of the individual CKM matrix elements $V^{\rm CKM}_{ab}$ are compared to test if unitarity conditions such as $\sum_a V^{\rm CKM}_{ab} V^{{\rm CKM}*}_{ac}= \delta_{bc}$ hold. When $b \neq c$, this relation leads to the celebrated unitarity triangles testing the unitarity of the CKM matrix. Thus, it is natural to consider whether such a successful probe may be applied to the PMNS matrix as well. 

However, the differences between the quark and neutrino sectors make this option somewhat inefficient when applied to the PMNS matrix. Indeed, while in hadronic weak decays the flavour of the up and down-type quarks can usually be tagged leading to a clean and precise measurement of the corresponding $V^{\rm CKM}_{ab}$ element, in the lepton sector only the charged fermions $\ell_\alpha$ are detected. As such, a direct measurement of the element $N_{\alpha i}$ is not generally possible, as it would require the identification of the neutrino mass eigenstate $\nu_i$. Instead, neutrino oscillation experiments usually tag the initial $\ell_\alpha$ and final $\ell_\beta$ and are thus sensitive to more convoluted combinations of matrix elements as in Eq.~\eqref{eq:normalised_probability}. This deteriorates the precision with which the individual $N_{\alpha i}$ elements can be recovered, especially if unitarity is not imposed, as this leads to too many free parameters for neutrino oscillation data to effectively constrain. 

Conversely, as mentioned above, matter effects or the zero distance effect provide very good sensitivity directly to the $\alpha$ parameters. In other words, they directly test the unitarity relation since $\sum_i N_{\beta i}^{} N_{\gamma i}^* \approx \delta_{\beta \gamma} -\alpha_{\beta \gamma} -\alpha^*_{\gamma \beta}$. Thus, with present data, our knowledge of the lepton mixing matrix is that it is close to unitary with higher accuracy than the uncertainty we typically have in its individual entries. As such, depicting this information through unitarity triangles (as for the CKM matrix) or measurements of the individual matrix elements may be somewhat misleading when used to transmit our uncertainty of the amount of non-unitarity presently allowed for the PMNS matrix. On the contrary, choosing a parametrization of these effects as in Eq.~\eqref{eq:eta_parametrisation} or Eq.~\eqref{eq:alpha_parametrisation}, more efficiently decouples the two sets of parameters. On the one hand, the unitary part of Eqs.~\eqref{eq:eta_parametrisation} and~\eqref{eq:alpha_parametrisation} controls the size of the PMNS matrix elements and is best determined through long-baseline experiments. On the other hand, the unitarity-violating parameters that are generally more efficiently probed via short-baseline searches or through regimes with strong matter effects, as well as via flavour and EWPO.

\subsection{Matter effects}
\label{subsec:matter}

Even though the flavour basis is not orthogonal, the mass basis is orthonormal. In it the Hamiltonian is Hermitian, so the evolution is unitary and can be computed in the usual manner. In particular, in the mass basis the Hamiltonian is given by~\cite{Antusch:2006vwa,Blennow:2016jkn}:

\begin{equation}
    H_{\rm vac}^m+H_{\rm mat}^m=\frac{1}{2E}
    \begin{pmatrix}
        0&&\\
        &\Delta m_{21}^2&\\
        &&\Delta m_{31}^2
    \end{pmatrix}
    + \sqrt{2} G_{F} N^t
    \begin{pmatrix}
        n_e-\tfrac12 n_n&&\\
        &-\tfrac12 n_n&\\
        &&-\tfrac12 n_n
    \end{pmatrix}
    N^*\,,
\end{equation}
where the second term corresponds to the matter potential in the mass basis assuming that matter is neutral, and containing contributions both from CC and NC interactions, which can be readily computed at leading order in terms of the non-unitarity parametrizations of Eqs.~\eqref{eq:eta_parametrisation} or~\eqref{eq:alpha_parametrisation}. In the isoscalar limit, in which the electron and neutron number densities $n_e$ and $n_n$ are equal\footnote{This is a fairly good approximation for the Earth, as the PREM model~\cite{Dziewonski:1981xy} fixes the average Earth neutron-to-proton ratio at $\left\langle Y_n^\oplus\right\rangle=1.051$.}, the matter potential reads: 
\begin{equation}
    H_{\rm mat,\oplus}^m= {\sqrt{2}} G_F n_e U^t
    \begin{pmatrix}
        1-\alpha_{ee}-\alpha_{\mu \mu} & \alpha_{\mu e}/2 & \alpha_{\tau e}/2\\
        \alpha_{\mu e}^*/2 & 0 & \alpha_{\tau \mu}/2\\
        \alpha_{\tau e}^*/2& \alpha_{\tau \mu}^*/2& \alpha_{\tau\tau}-\alpha_{\mu \mu}
    \end{pmatrix}
    U^*+\mathcal O(\alpha^2)\,,
    \label{eq:matter_potential_not_GF_corrected}
\end{equation}
where we have subtracted $\alpha_{\mu\mu}$ from the diagonal entries, as a term proportional to the identity in the Hamiltonian does not contribute to oscillations. Equivalently, in the $\eta$-parametrization, the matter potential reads:

\begin{equation}
    H_{\rm mat,\oplus}^m=\sqrt{2} G_F n_e V^t
    \begin{pmatrix}
        1-\eta_{ee}-\eta_{\mu \mu} & 0 & 0\\
        0 & 0 & \eta_{\tau \mu}\\
        0& \eta_{\tau \mu}^*& \eta_{\tau\tau}-\eta_{\mu \mu}
    \end{pmatrix}
    V^*+\mathcal O(\eta^2)\,.
        \label{eq:matter_potential_not_GF_corrected_mu}
\end{equation}
Notice the very different structure of Eqs.~\eqref{eq:matter_potential_not_GF_corrected} and~\eqref{eq:matter_potential_not_GF_corrected_mu}. This is due to cancellations that take place in the $\mu-e$ and $\tau-e$ entries when the isoscalar limit is taken for the $\eta$-parametrization. The solution to this apparent inconsistency, that may also lead to misconceptions, is again the fact that that $U$ and $V$ are not the same matrix~\cite{Blennow:2016jkn}.

Another important observation often overlooked in the literature is that, in the presence of a non-unitary leptonic mixing matrix, the Fermi constant that enters in the above equations does not coincide with the quantity that is experimentally extracted from muon decay. Indeed, as it proceeds through CC interactions involving neutrinos, non-unitarity will also affect the muon decay rate $\Gamma_\mu$ and modify the extraction of the Fermi constant:
\begin{equation}
    \Gamma_\mu=\frac{G_F^2m_\mu^5}{192\pi^3}\sum_{i=1}^3\left\lvert N_{\mu i}\right\rvert^2 \sum_{j=1}^3\left\lvert N_{ej}\right\rvert^2 \simeq \frac{G_F^2m_\mu^5}{192\pi^3}\left(1-2\alpha_{ee}-2\alpha_{\mu\mu}\right)\equiv \frac{G_\mu^2m_\mu^5}{192\pi^3}\,.
\end{equation}
As a consequence, the value of the Fermi constant extracted form muon decay ($G_\mu$) is related to the actual Fermi constant ($G_F$) entering, for example, electroweak precision observables and, in particular, the above matter potential, in the following way:
\begin{equation}
    G_F \simeq G_\mu(1+\alpha_{ee}+\alpha_{\mu\mu})\,.
\end{equation}
Upon substituting the above expression in Eq.~\eqref{eq:matter_potential_not_GF_corrected}, there is an exact cancellation~\cite{Aloni:2022ebm} at leading order:
\begin{equation}
    H_{\rm mat,\oplus}^m= {\sqrt{2}} G_\mu n_e U^t
    \begin{pmatrix}
        1 & \alpha_{\mu e}/2 & \alpha_{\tau e}/2 \\
        \alpha_{\mu e}^*/2 & 0 & \alpha_{\tau \mu}/2\\
        \alpha_{\tau e}^*/2& \alpha^*_{\tau \mu}/2& \alpha_{\tau\tau}-\alpha_{\mu \mu}
    \end{pmatrix}
    U^*+\mathcal O(\alpha^2)\,,
    \label{eq:matter_potential_GF_corrected}
\end{equation}
which removes all sensitivity to $\alpha_{ee}+\alpha_{\mu\mu}$ in neutrino propagation\footnote{Alternatively, it could be possible to use as input a value of $G_F$ not affected by non-unitarity, for instance extracted from the value of the $W$ mass $M_W$. Nevertheless, much better sensitivity is obtained by comparing this value with $G_\mu$ (see {\it e.g.}~Ref.~\cite{Blennow:2023mqx}), given the much higher accuracy as compared to measurements of the neutrino matter effects. } through the Earth in the isoscalar limit. Beyond the $n_e=n_n$ limit some sensitivity will remain, although at cuadratic order or suppressed at linear order by $\left\langle Y_{n}^{\oplus}\right\rangle-1=0.051$. We thus conclude that the main sensitivity that neutrino propagation in the Earth has to diagonal non-unitarity parameters is the combination $\alpha_{\tau\tau}-\alpha_{\mu\mu}$.

\subsection{Neutral-Current detection}

There are samples of neutrino events measured through neutral current interactions (NC) that can be useful to constrain new physics and in general will also be sensitive to non-unitarity. 
When the neutrino detection is performed via NC, the outgoing neutrino is not measured and, as such, the number of events would be given by the incoherent sum of the different possible independent final states. Thus, naively, a sum over all the flavour states would provide the desired event rate and hence be proportional to the ``experimental probability''. In fact, this has been the approach often adopted:
\begin{equation}
\tilde{P}_{\nu_\alpha \to \nu_\mathit{NC}}^\mathit{\rm naive}(L) \equiv \sum_\beta P_{\alpha\beta} = \frac{1}{\mathcal{N}}\sum\limits_{\beta} \Big|\sum\limits_{ i} N_{\alpha i}^* N_{\beta i}^{}  e^{-i\frac{m_i^2 L}{2E}} \Big|^2 = \frac{1}{\mathcal{N}}\sum_{\beta, i, j} N_{\alpha i}^* N_{\beta i} N_{\alpha j} N_{\beta j}^* e^{i \Delta m^2_{ji} L/2E}\, ,
\label{eq:naiveNCprob}
\end{equation}
where, for simplicity, we have assumed vacuum propagation and $\mathcal{N}$ is the necessary normalization factor which will depend on how the neutrino flux and cross section is estimated, as discussed in Section~\ref{sec:experimental_probabilities}.

However, as emphasized in Section~\ref{sec:theoretical_probabilities}, the different neutrino flavour eigenstates are \textit{not} mutually exclusive since they are not orthogonal. As such, the incoherent sum in Eq.~\eqref{eq:naiveNCprob} is not correct and it should rather be performed in the orthonormal mass basis, which truly represent independent outcomes (see e.g. the discussion in Ref.~\cite{Coloma:2022umy}). Moreover, it is also important to notice that, in presence of non-unitary mixing, NC interactions are not proportional to $\delta_{ij}$ in the mass basis but instead to $N^\dagger N$, as given by Eq.~\eqref{eq:lagr}. As such, and in complete analogy to the $N_{\alpha i}^*$ factor from the production of a $|\nu_\alpha \rangle$ via CC interactions, the NC detection rate of a $|\nu_\alpha \rangle$ is given by~\cite{Dutta:2019hmb}:
\begin{align}
\label{eq:NCprob}
\tilde{P}_{\nu_\alpha \to \nu_{\mathit{NC}}}(L) &=  
\frac{1}{\mathcal{N}} \sum_k \left|N^* \mathcal{S} \, (N^\dagger N)^*\right|^2_{\alpha k}  =
\frac{1}{\mathcal{N}}
\sum\limits_{k} \Big|\sum\limits_{\beta, i} N_{\alpha i}^{*}e^{-i\frac{m_i^2 L}{2E}} N_{\beta i} N_{\beta k}^{*}  \Big|^2 \nonumber\\ & = \frac{1}{\mathcal{N}}
\sum\limits_{k, \beta, i, \gamma, j} 
N_{\alpha i}^* N_{\beta i}^{} N_{\beta k}^*N_{\alpha j}^{} N_{\gamma j}^* N_{\gamma k}^{} \, e^{i\frac{\Delta m^2_{ji} L}{2E}}
\,,
\end{align}
where $S$ denotes the evolution operator. Using that 
\begin{equation}
\sum_k N_{\gamma k}^{} N_{\beta k}^* = \delta_{\beta \gamma} - \alpha_{\beta \gamma}^* - \alpha_{\gamma \beta } + \mathcal O(\alpha^2)\,,
\end{equation}
in Eq.~\eqref{eq:NCprob} we obtain:
\begin{align}
\label{eq:NCprobcorrection}
\tilde{P}_{\nu_\alpha \to \nu_\mathit{NC}}(L) & =\tilde{P}_{\nu_\alpha \to \nu_\mathit{NC}}^\mathit{naive}(L)  
 - \frac{1}{\mathcal{N}}\Bigg\{\sum_{\beta,\gamma, i,j} N_{\alpha i}^* N_{\beta i}^{} N_{\alpha j}^{} N_{\gamma j}^* \alpha_{\beta \gamma}^*  e^{i \Delta m^2_{ji} L/2E}\nonumber\\
& - N_{\alpha i}^* N_{\beta i}^{} N_{\alpha j}^{} N_{\gamma j}^* \alpha_{\gamma \beta}  e^{i \Delta m^2_{ji} L/2E} + \mathcal O(\alpha^2) \Bigg\}\,.
\end{align}
Hence, there are leading order corrections to the naive approach of adding incoherently over the flavour states often adopted that must be considered to obtain the correct result.

A further subtlety is that, for NC detection, it does not seem possible to define a useful ``theoretical probability'' as in Section~\ref{sec:theoretical_probabilities} representing a true probability, that is, an overlap between normalized states. Indeed, since NC detection goes with the elements of $(N^\dagger N)_{ij} \neq \delta_{ij}$, each mass eigenstate contribution to the NC detection is weighted by a different factor. Thus, the NC detection process could be written as a weighted incoherent sum of actual probabilities, but not as a probability itself. Nevertheless, the corresponding ``experimental probability'', as in Section~\ref{sec:experimental_probabilities}, can be defined normalizing Eq.~\eqref{eq:NCprobcorrection} properly through $\mathcal{N}$ according to how the NC even rate is estimated in the corresponding experiment. For example, if the expectation is calibrated by a NC near detector measurement of the $|\nu_\alpha\rangle$ flux, the normalization factor would read:
\begin{equation}
    \mathcal{N}=\sum\limits_{k} \Big|\sum\limits_{\beta, i} N_{\alpha i}^* N_{\beta i}^{} N_{\beta k}^* \Big|^2\,,
\end{equation}
and the correctly normalized ``experimental probability'' would be given by:
\begin{equation}
\label{eq:expNCprob}
\tilde{P}_{\nu_\alpha \to \nu_{\mathit{NC}}}(L) = \frac{\sum\limits_{k} \Big|\sum\limits_{\beta, i} N_{\alpha i}^* N_{\beta i}^{} N_{\beta k}^* e^{-i\frac{m_i^2 L}{2E}} \Big|^2}{\sum\limits_{k} \big|\sum\limits_{\beta, i} N_{\alpha i}^* N_{\beta i}^{} N_{\beta k}^* \big|^2}\,.
\end{equation}

%%%%%%%%%%%%%%%%%%%%%%%%%%%%%%%%%%%%%%%%%%%%%%%%%%%%%%%%%%%%%%%
\section{Applicability of the different non-unitarity bounds}
\label{sec:bounds}

Given the fact that the neutrino oscillation phenomenon is sensitive to the elements of the non-unitary leptonic mixing matrix, it could be expected to provide its most direct probe. Nevertheless, the Lagrangian shown in Eq.~\eqref{eq:lagr} has other very important and sensitive phenomenological consequences. Indeed, since both CC and NC neutrino interactions are modified, very well-measured processes from which fundamental parameters are extracted are affected. This is the case, as already outlined, of $G_F$ from $\mu$ decay, whose amplitude has two insertions of $N_{\alpha i}$. Similarly, $\beta$, $\pi$ and $\tau$ decays are affected. This means that electroweak precision observables and lepton universality tests can provide exquisite sensitivity to unitarity deviations of $N$. Moreover, flavour-violating processes can also provide relevant constraints since the GIM mechanism~\cite{Glashow:1970gm} suppressing them in the SM stems from the unitarity of the mixing matrix, which is now lost. These sets of observables have been used to derive very stringent constraints on how large unitarity deviations are presently allowed to be. For instance, the most recent global fit~\cite{Blennow:2023mqx} constrains most elements of $\alpha$ to be below the $10^{-3}$ level at $2\sigma$, beyond the sensitivity of present or near future neutrino oscillation experiments. It is then interesting to consider the applicability and potential complementarity between the two sets of constraints.

It has become common to refer to bounds stemming from flavour and EWPO as ``charged lepton'' or ``indirect'' bounds and those from neutrino oscillations as ``neutrino'' or ``direct'' bounds, somehow implying that the latter are more robust, even if they are in principle weaker. However, in both cases the sensitivity to the mixing matrix and its possible unitarity deviations mainly stems from processes involving CC interactions, thus always involving both a neutrino and a charged lepton. Their characterization as ``charged lepton'' or ``neutrino'' bounds is therefore spurious. In fact, there is a large overlap between the processes involved in the constraints from flavour and EWPO and those responsible for neutrino production and detection in neutrino oscillation experiments. For example, the same $\pi$ decays through which the neutrino fluxes for many neutrino oscillation experiments are produced, are also the source of the strong constraints originating from weak lepton universality tests given by its ratios.
Both involve neutrinos and charged leptons and are, in principle, equally applicable and robust, since they are based in the same underlying process. Similar arguments can be applied to neutrino detection via inverse $\beta$ decay and the constraints coming from $\beta$ or $\mu$ decays in EWPO. 

In a more general framework, one could consider introducing appropriate 4-fermion operators precisely canceling all the effects of a non-unitary mixing matrix in EWPO. However, these 4-fermion operators would also induce neutrino non-standard interactions (NSI) also canceling the effect of non-unitarity in neutrino oscillations, as expected. Indeed, as discussed in Ref.~\cite{Blennow:2016jkn}, by integrating out the $W$ and the $Z$, all NU effects can be described as a particular realization of the NSI formalism involving both NC and CC NSI with particular correlations among their coefficients. Thus, the two sets of observables are generally equally robust. 

However, some of the EWPO might be affected by new physics contributions that do not affect neutrino oscillations similarly. For instance, one of the strongest constraints comes from direct measurements of $M_W$ when comparing with $G_F$ as extracted from $\mu$ decay. If $\mu$ decay is modified, for instance via a dimension-6 4-fermion operator, it could cancel the effect of non-unitarity in $\mu$ decay and avoid this constraint, while neutrino oscillations not involving production from $\mu$ decay would however not be affected. Similarly, new physics affecting the $T$ parameter may help to avoid this constrain without significantly affecting neutrino oscillations. The constraints from lepton flavour universality tests are, on the other hand, more difficult to avoid without losing as well the sensitivity in neutrino oscillation experiments. Particularly those stemming from $\pi$ decay. Moreover, apart from the measurement of $M_W$, precise measurements of $\sin \theta_W$ at different energies as well as $\beta$ decays also provide strong constraints and would all need to be avoided by different {\it ad hoc} new physics, which seems rather challenging. Thus, while it is technically possible to avoid some of the flavour and EWPO constraints without affecting neutrino oscillations, it is, in general, not possible to avoid all of them without also losing the effect in the oscillation phenomenon itself. 

\subsection*{Non-unitarity from light sterile neutrinos}

An interesting way of evading all flavour and EWPO constraints on non-unitarity is when it is originated by extra sterile neutrinos which are light enough to be produced in all relevant processes down to $\beta$ decays. In this scenario, unitarity is restored in all flavour and EWPO, as all the mass eigenstates are kinematically accessible. Conversely, these neutrinos would be produced and propagate together with the 3 standard light neutrinos, participating of the oscillation phenomenon. Thus, an effect is still present in neutrino oscillation experiments even if flavour and EWPO constraints are lost. 

If the new mass eigenstates are light enough, they will lead to new oscillation frequencies that might be probed for at neutrino oscillation experiments looking for sterile neutrinos. Conversely, if the new mass eigenstates are characterized by masses that lead to large $\Delta m^2 L/E$, their oscillations will average out at the detector or lose coherence. Interestingly, as discussed in Refs.~\cite{Blennow:2016jkn,Fong:2016yyh}, the phenomenology of sterile neutrinos in this averaged-out regime and that of non-unitarity is the same except for a constant term that is generally subleading. Moreover, the normalizations discussed in Section~\ref{sec:experimental_probabilities} are typically not present in this scenario, as long as the sterile oscillations have not developed yet at the near detector or in the data used to normalize the flux or cross section. This may lead to an enhancement of the sensitivity compared to the ``standard'' scenario covered in the previous sections, as discussed in Ref.~\cite{Blennow:2016jkn}. While this scenario does not fully correspond with the definition of non-unitarity introduced in Section~\ref{sec:nouni}, the phenomenological impact in the neutrino oscillation phenomenon is almost the same and non-unitarity constraints from neutrino oscillation searches largely apply to this regime, while flavour and EWPO do not. Indeed it is not uncommon to find analyses in the literature referring to this regime also as non-unitarity, even though it may more naturally and accurately be described by the large $\Delta m^2$  limit of the sterile neutrino framework~\cite{Acero:2022wqg}. 

The $3+1$ sterile neutrino analyses in short baseline experiments are typically performed as a function of the mixing between the active and mostly sterile states $\mathcal{U}_{\alpha 4}$ or the effective mixing angles $\vartheta_{\alpha \beta}$:
\begin{align}
\tilde{P}_{\nu_\alpha\to\nu_\beta}(L)& \equiv \sin^2 2 \vartheta_{\alpha \beta} \sin^2 \left(\frac{\Delta m^2_{41} L}{4E} \right) \,,\\
\tilde{P}_{\nu_\alpha\to\nu_\alpha}(L)& \equiv 1-\sin^2 2 \vartheta_{\alpha \alpha} \sin^2 \left(\frac{\Delta m^2_{41} L}{4E} \right) \,.
\end{align}
However, as we have already mentioned, it can also be performed in terms of the $\alpha$ parameters in the averaged-out sterile neutrino scenario. The connection between both parameterizations can be easily derived by considering the unitarity of the full mixing matrix $\mathcal{U}\mathcal{U}^\dagger=I$. In the common $3+1$ case we have\footnote{Notice that the relations in Eqs.~\eqref{eq:alphavsU1}-\eqref{eq:alphavsU2} are only valid for small mixing angles.}~\cite{Blennow:2016jkn,Coloma:2021uhq}
\begin{align}
\alpha_{\beta\beta}&\simeq\frac{1}{2}|\mathcal{U}_{\beta 4}|^2 \simeq \frac{1}{2}\sin^2 \theta_{\beta 4} \simeq \frac{1}{2}\sin^2\vartheta_{\beta\beta}\,,\label{eq:alphavsU1}\\
|\alpha_{\gamma\beta}|&\simeq |\mathcal{U}_{\gamma 4}\mathcal{U}_{\beta 4}^*| \simeq |\sin \theta_{\gamma 4}||\sin \theta_{\beta 4}| \simeq \frac{1}{2}|\sin2\vartheta_{\gamma\beta}|\,,
\label{eq:alphavsU2}
\end{align}
with $\gamma > \beta$.

\begin{table}[t!]
    \centering
    \begin{tabularx}{0.9\textwidth}{|c| Y  Y | Y Y|}
    \hline
         &\multicolumn{2}{c|}{}&\multicolumn{2}{c|}{}\\[-2.9ex]
    \multirow{3}{*}{\bf 90\%\,CL}&\multicolumn{2}{c}{\bf Averaged $\boldsymbol{\nu}$ Oscillations } 
    &\multicolumn{2}{c}{\bf Flavour \& EWPO~\cite{Blennow:2023mqx} }\\[0ex]  
        & \multicolumn{2}{c|}{$\mathbf{m>10}$~{\bf eV}}  & \multicolumn{2}{c|}{$\mathbf{m >M_Z}$} \\[0ex]
     \cline{2-5}&&&&\\[-2.2ex]
    & {\bf Direct } & {\bf Schwarz} & {\bf Direct} & {\bf Schwarz}\\
    \hline
    &&&&\\[-2.ex]
    
    $\alpha_{ee}$ & $8.4\times 10^{-3}$~\cite{Goldhagen:2021kxe}&-&$1.2\times 10^{-3}$&-\\[1ex]  
    &&&&\\[-2.ex]
    
    $\alpha_{\mu\mu}$&$1.2\times 10^{-2}$~\cite{MINOS:2020iqj}&-&$8.6\times 10^{-5}$&-\\[1ex]
    &&&&\\[-2.ex]
    
    $\alpha_{\tau\tau}$&$2.9\times10^{-2}$~\cite{IceCube:2024dlz}&-&$6.0\times10^{-4}$&-\\[1ex]
    &&&&\\[-2.ex]
   
    $\left\lvert\alpha_{\mu e}\right\rvert$&$1.8\times 10^{-2}$~\cite{NOMAD:2003mqg}&$1.4\times 10^{-2}$&$1.9\times 10^{-5}$&$5.4\times 10^{-4}$\\[1ex]
    &&&&\\[-2.ex]
   
    $\left\lvert\alpha_{\tau e}\right\rvert$&$6.1\times10^{-2}$~\cite{NOMAD:2001xxt}&$2.2\times 10^{-2}$&$6.2\times10^{-3}$&$1.5\times 10^{-3}$\\[1ex]\
    &&&&\\[-2.ex]
   
    $\left\lvert\alpha_{\tau\mu}\right\rvert$&$9.1\times10^{-3}$~\cite{NOMAD:2001xxt}&$2.6\times10^{-2}$&$6.9\times 10^{-3}$&$1.5\times 10^{-4}$\\[1ex]\hline
    \end{tabularx}
    \caption{Constraints on the non-unitarity parameters $\alpha_{\beta \gamma}$ from neutrino oscillation searches (left columns) and flavour and EWPO (right columns). For the oscillation bounds we report constraints that apply for sterile neutrino masses in the averaged-out regime with $\Delta m^2 >100$~eV$^2$. For lighter masses, stronger constraints exploiting the new oscillation frequencies generally apply. All flavour and EWPO bounds apply for sterile neutrino masses at least $m>M_Z$. For lighter sterile neutrinos down to $m \sim 1$~MeV stronger constraints from direct searches may apply. See main text for further details. }
    \label{tab:my_label}
\end{table}

In Table~\ref{tab:my_label} we compare the bounds on the non-unitarity parameters $\alpha$ from EWPO (right column) and neutrino oscillation experiments when assuming sterile neutrinos in the averaged-out regime of their new $\Delta m^2$ (left column). The latter, compiled and updated for this work, mainly come from short baseline experiments, therefore probing the ``zero distance effect'', or searches exploiting sizable matter effects, which are the most sensitive probes, as discussed in the previous sections. For all constraints displayed, the bounds apply from at least $\Delta m^2 > 100$~eV$^2$. Nevertheless, several of them also apply for smaller values and, in any event, for $\Delta m^2$ values below the averaged-out regime, generally stronger constraints apply since sensitivity can improve exploiting also the shape information. Constraints are particularly strong around the $\Delta m^2 = 1$~eV$^2$ region. Indeed, several dedicated experiments have probed the long-standing short-baseline neutrino anomalies which could be interpreted as signal but are also in strong tension with other observations~\cite{Dentler:2018sju}. Therefore, we do not include the short-baseline neutrino anomalous results in the set of data used to derive the constraints. Regarding the flavour and EWPO, all constraints apply at least for neutrino masses above the mass of the $Z$. While several of the constraints would also apply down to the $\mu$ mass, generally stronger bounds are also in place from direct searches for sterile neutrinos. These go from $\sim 100$~GeV through collider searches down to $\sim 1$~MeV from production through meson decays or even to $\sim 1$~keV for $\alpha_{ee}$ from searches at $\beta$ decays (see for example Ref.~\cite{Atre:2009rg,Fernandez-Martinez:2023phj}).

For the flavour off-diagonal elements, besides the {\it direct} constraints already mentioned, we also provide those derived from the Schwarz inequality 
\begin{equation}
|\alpha_{\beta \gamma}| \leq 2 \sqrt{\alpha_{\beta \beta} \alpha_{\gamma \gamma}}\,.
\end{equation} 
This relation follows when the coefficient of the $d=6$ operator of Eq.~\eqref{eq:d6op} $\eta$ is positive (or negative) definite. This is the case of the type-I Seesaw mechanism, which is the only UV completion that leads only to the $d=6$ operator of Eq.~\eqref{eq:d6op}. Other sources of non-unitarity will generally also contribute to other operators usually affecting the charged lepton sector and therefore more strongly constrained. Nevertheless, it is technically possible to combine several UV completions and tune their respective couplings to cancel other contributions and avoid the applicability of the Schwarz inequality. Ref.~\cite{Coutinho:2019aiy} briefly describes the level of complexity required to achieve this through a particular example. In such a case, only the {\it direct} bounds would apply.

%%%%%%%%%%%%%%%%%%%%%%%%%%%%%%%%%%%%%%%%%%

To summarize, non-unitarity of the lepton mixing matrix may be induced by many different generic high-energy extensions of the SM if, upon integrating out their new heavy degrees of freedom, a contribution to the $d=6$ operator of Eq.~\eqref{eq:d6op} is obtained. These scenarios are best probed by flavour and EWPO and the role of neutrino oscillation constraints is generally subleading. Further new physics beyond the SM is necessary to explain the origin of neutrino masses and mixings, and the simplest and most natural extension is to add right-handed neutrinos to the SM particle content. If these neutrinos are too heavy to be directly produced, they can still be probed for indirectly through the non-unitarity they induce and the strongest constraints are from flavour and EWPO. Conversely, when they are light enough to be produced and participate together with the three standard neutrinos in the oscillation phenomenon, neutrino oscillation experiments can be the main window to test their existence, as most other constraints are lost. In this sense, the two sets of constraints on non-unitarity can be viewed as complementary, as they are probing two different parts of the parameter space of this simplest SM extension.

\section{Conclusions}
\label{sec:conc}

We have highlighted a number of subtle issues that arise when considering neutrino oscillations in presence of a non-unitary leptonic mixing matrix. In particular, after introducing the framework and possible sources of non-unitarity we have discussed that:

\begin{itemize}

\item In general, oscillation probabilities do not match with the experimental observable and we proposed two possible definitions. The first is a ``theoretical probability'' which is a true probability but may not correspond to the measured quantity. The second is the ``experimental probability'', which more directly corresponds to what is determined at a given experiment, but may not represent an actual probability. While the two may coincide, this is not always the case in presence of non-unitarity.

\item In presence of a non-unitary mixing matrix, the flavour eigenstates are not orthogonal and hence not mutually exclusive. As such, even the ``theoretical probabilities'' describing transitions to different flavours do not need to add up to 1, which has sourced some misunderstandings in the literature. Similarly, when considering detection via NC interactions, the appropriate basis is the mass basis, since the states are orthogonal and mutually exclusive. Naively performing the computation in the flavour basis may lead to incorrect results. 

\item The non-orthogonality of the flavour eigenstates also leads to flavour transitions at zero distance. This ``zero distance effect'' has often been considered in disappearance channels as a source of strong constraints, given its linear dependence with the non-unitarity parameters. However, when properly normalizing the oscillation probabilities, the ``zero distance effect'' cancels in disappearance channels and care must be taken. Similarly, also at longer baselines, some sensitivity to non-unitarity is lost in disappearance channels.  

\item It is convenient to parametrize a non-unitary mixing matrix through a unitary one and some additional (small) parameters that encode the unitarity deviations. A common misconception is to use as input for the unitary part the outcome of global fits to oscillation data that assume unitarity in their analysis. This neglects the non-unitary corrections that would also affect the experiments analyzed in the global fit. This procedure leads to different and inconsistent results depending on the parametrization employed, as the non-unitary corrections neglected in the unitary global fit are of leading order in the non-unitary parameters. Some parametrizations, like the popular lower triangular introduced in Eq.~\eqref{eq:alpha_parametrisation}, can mitigate its impact, making non-unitary corrections essentially absent in the extraction of $\theta_{13}$ and subleading for $\theta_{23}$. Unfortunately, matter effects in solar neutrino experiments do make an extraction of $\theta_{12}$ free from non-unitary corrections challenging.

\item Unlike for the CKM, the best sensitivity to non-unitarity effects in the lepton sector \emph{does not arise} from a precise determination of the individual elements in the PMNS matrix in oscillation data. Instead, the unitarity-violating parameters
are generally more efficiently probed via short-baseline searches or through regimes with strong matter effects, as well as via flavour and EWPO. Therefore, depicting our knowledge of the unitarity of the PMNS matrix through unitarity triangles is not as convenient. 

\item  A non-unitary mixing matrix also affects the rates of weak processes involving neutrinos such as $\mu$, $\pi$, $\beta$ or $\tau$ decays. Given its impact in weak decays, flavour and EWPO also provide constraints on the unitarity of the lepton mixing matrix which are generally stronger than the sensitivity found for neutrino oscillation experiments. While it is sometimes argued that the latter constitute more ``direct'' or ``robust`` constraints, this is not the case, as the sensitivity is in both cases provided mainly by the same underlying processes. In Section~\ref{sec:bounds} we have updated and compared both sets of constraints.

\item The usual extraction of $G_F$ from $\mu$ decay is affected and this indirect dependence on the non-unitarity parameters should also be taken into account consistently when considering the impact of matter effects in oscillation experiments. This also cancels some sensitivity. 

\end{itemize}

 Finally, an interesting regime related to non-unitarity is neutrino oscillations in presence of sterile neutrinos that are light enough to be produced together with the three standard neutrinos but too heavy for their oscillations to be seen at the far detector (oscillations average out or decohere). In this regime, the phenomenology is largely the same as in presence of a non-unitary mixing matrix but the normalization of the ``experimental probability'' may not be present, leading to an increased sensitivity and recovering ``zero distance effects'' in disappearance channels. Moreover, the stronger constraints from flavour and EWPO do not apply, making neutrino oscillation experiments the best window to probe this scenario.

%\medskip

\paragraph{Acknowledgments.} 
This project has received support from the European Union’s Horizon 2020 research and innovation programme under the Marie Skłodowska-Curie grant agreement No~860881-HIDDeN and No 101086085 - ASYMMETRY, and from the Spanish Research Agency (Agencia Estatal de Investigaci\'on) through the Grant IFIC Centro de Excelencia Severo Ochoa No CEX2023-001292-S, Grant PID2020-113644GB-I00, Grant IFT Centro de Excelencia Severo Ochoa No CEX2020-001007-S and Grant PID2022-137127NB-I00 funded by MCIN/AEI/10.13039/501100011033.  
The work of DNT was supported by the Spanish MIU through the National Program FPU (grant number FPU20/05333). JLP also acknowledges financial support from Generalitat Valenciana through the plan GenT program (CIESGT2024-008), from the Spanish Research Agency (Agencia Estatal de Investigaci\'on) through grant CNS2022-136013 funded by MICIU/AEI/10.13039/501100011033 and by “European Union NextGenerationEU/PRTR'', and from the MCIU with funding from the European Union NextGenerationEU (PRTR-C17.I01) and Generalitat Valenciana (ASFAE/2022/020). PC acknowledges support from Grant PID2022-142545NB-C21 funded by MCIN/AEI/10.13039/501100011033/ FEDER, UE, from the Spanish Research Agency through the grant CNS2023-145338 funded by MCIN/AEI/10.13039/501100011033 and by “European Union NextGenerationEU/PRTR''. She is also supported by grant RYC2018-024240-I, funded by MCIN/AEI/10.13039/501100011033 and by ``ESF Investing in your future''.

%%%%%%%%%%%%%%%%%%%%%%%%%%%%%%%%%%%%%%%%%
%\appendix

%%%%%%%%%%%%%%%%%%%%%%%%%%%%%%%%%%%%%%%%

\bibliographystyle{JHEP} 
\bibliography{biblio}% Produces the bibliography via BibTeX.

\providecommand{\href}[2]{#2}\begingroup\raggedright\begin{thebibliography}{100}

\bibitem{Glashow:1970gm}
S.~L. Glashow, J.~Iliopoulos, and L.~Maiani, {\it {Weak Interactions with
  Lepton-Hadron Symmetry}},  {\em Phys. Rev. D} {\bf 2} (1970) 1285--1292.

\bibitem{Esteban:2024eli}
I.~Esteban, M.~C. Gonzalez-Garcia, M.~Maltoni, I.~Martinez-Soler, J.~a.~P.
  Pinheiro, and T.~Schwetz, {\it {NuFit-6.0: updated global analysis of
  three-flavor neutrino oscillations}},  {\em JHEP} {\bf 12} (2024) 216,
  [\href{http://arxiv.org/abs/2410.05380}{{\tt arXiv:2410.05380}}].

\bibitem{Capozzi:2021fjo}
F.~Capozzi, E.~Di~Valentino, E.~Lisi, A.~Marrone, A.~Melchiorri, and
  A.~Palazzo, {\it {Unfinished fabric of the three neutrino paradigm}},  {\em
  Phys. Rev. D} {\bf 104} (2021), no.~8 083031,
  [\href{http://arxiv.org/abs/2107.00532}{{\tt arXiv:2107.00532}}].

\bibitem{deSalas:2020pgw}
P.~F. de~Salas, D.~V. Forero, S.~Gariazzo, P.~Mart\'\i{}nez-Mirav\'e, O.~Mena,
  C.~A. Ternes, M.~T\'ortola, and J.~W.~F. Valle, {\it {2020 global
  reassessment of the neutrino oscillation picture}},  {\em JHEP} {\bf 02}
  (2021) 071, [\href{http://arxiv.org/abs/2006.11237}{{\tt arXiv:2006.11237}}].

\bibitem{Minkowski:1977sc}
P.~Minkowski, {\it {$\mu \to e\gamma$ at a Rate of One Out of $10^{9}$ Muon
  Decays?}},  {\em Phys. Lett. B} {\bf 67} (1977) 421--428.

\bibitem{Mohapatra:1979ia}
R.~N. Mohapatra and G.~Senjanovic, {\it {Neutrino Mass and Spontaneous Parity
  Nonconservation}},  {\em Phys. Rev. Lett.} {\bf 44} (1980) 912.

\bibitem{Yanagida:1979as}
T.~Yanagida, {\it {Horizontal gauge symmetry and masses of neutrinos}},  {\em
  Conf. Proc. C} {\bf 7902131} (1979) 95--99.

\bibitem{Gell-Mann:1979vob}
M.~Gell-Mann, P.~Ramond, and R.~Slansky, {\it {Complex Spinors and Unified
  Theories}},  {\em Conf. Proc. C} {\bf 790927} (1979) 315--321,
  [\href{http://arxiv.org/abs/1306.4669}{{\tt arXiv:1306.4669}}].

\bibitem{Broncano:2002rw}
A.~Broncano, M.~B. Gavela, and E.~E. Jenkins, {\it {The Effective Lagrangian
  for the seesaw model of neutrino mass and leptogenesis}},  {\em Phys. Lett.
  B} {\bf 552} (2003) 177--184,
  [\href{http://arxiv.org/abs/hep-ph/0210271}{{\tt hep-ph/0210271}}]. [Erratum:
  Phys.Lett.B 636, 332 (2006)].

\bibitem{Petcov:1976ff}
S.~T. Petcov, {\it {The Processes $\mu \rightarrow e + \gamma, \mu \rightarrow
  e + \overline{e}, \nu' \rightarrow \nu + \gamma$ in the Weinberg-Salam Model
  with Neutrino Mixing}},  {\em Sov. J. Nucl. Phys.} {\bf 25} (1977) 340.
  [Erratum: Sov.J.Nucl.Phys. 25, 698 (1977), Erratum: Yad.Fiz. 25, 1336
  (1977)].

\bibitem{Bilenky:1977du}
S.~M. Bilenky, S.~T. Petcov, and B.~Pontecorvo, {\it {Lepton Mixing, mu
  --\ensuremath{>} e + gamma Decay and Neutrino Oscillations}},  {\em Phys.
  Lett. B} {\bf 67} (1977) 309.

\bibitem{Cheng:1977vn}
T.-P. Cheng and L.-F. Li, {\it {Muon Number Nonconservation in Gauge
  Theories}},  {\em Stud. Nat. Sci.} {\bf 12} (1977) 659--681.

\bibitem{Marciano:1977wx}
W.~J. Marciano and A.~I. Sanda, {\it {Exotic Decays of the Muon and Heavy
  Leptons in Gauge Theories}},  {\em Phys. Lett. B} {\bf 67} (1977) 303--305.

\bibitem{Lee:1977qz}
B.~W. Lee, S.~Pakvasa, R.~E. Shrock, and H.~Sugawara, {\it {Muon and Electron
  Number Nonconservation in a V-A Gauge Model}},  {\em Phys. Rev. Lett.} {\bf
  38} (1977) 937. [Erratum: Phys.Rev.Lett. 38, 1230 (1977)].

\bibitem{Lee:1977tib}
B.~W. Lee and R.~E. Shrock, {\it {Natural Suppression of Symmetry Violation in
  Gauge Theories: Muon - Lepton and Electron Lepton Number Nonconservation}},
  {\em Phys. Rev. D} {\bf 16} (1977) 1444.

\bibitem{Shrock:1980vy}
R.~E. Shrock, {\it {New Tests For, and Bounds On, Neutrino Masses and Lepton
  Mixing}},  {\em Phys. Lett. B} {\bf 96} (1980) 159--164.

\bibitem{Schechter:1980gr}
J.~Schechter and J.~W.~F. Valle, {\it {Neutrino Masses in SU(2) x U(1)
  Theories}},  {\em Phys. Rev. D} {\bf 22} (1980) 2227.

\bibitem{Shrock:1980ct}
R.~E. Shrock, {\it {General Theory of Weak Leptonic and Semileptonic Decays. 1.
  Leptonic Pseudoscalar Meson Decays, with Associated Tests For, and Bounds on,
  Neutrino Masses and Lepton Mixing}},  {\em Phys. Rev. D} {\bf 24} (1981)
  1232.

\bibitem{Shrock:1981wq}
R.~E. Shrock, {\it {General Theory of Weak Processes Involving Neutrinos. 2.
  Pure Leptonic Decays}},  {\em Phys. Rev. D} {\bf 24} (1981) 1275.

\bibitem{Langacker:1988ur}
P.~Langacker and D.~London, {\it {Mixing Between Ordinary and Exotic
  Fermions}},  {\em Phys. Rev. D} {\bf 38} (1988) 886.

\bibitem{Pilaftsis:1992st}
A.~Pilaftsis, {\it {Lepton flavor nonconservation in H0 decays}},  {\em Phys.
  Lett. B} {\bf 285} (1992) 68--74.

\bibitem{Ilakovac:1994kj}
A.~Ilakovac and A.~Pilaftsis, {\it {Flavor violating charged lepton decays in
  seesaw-type models}},  {\em Nucl. Phys. B} {\bf 437} (1995) 491,
  [\href{http://arxiv.org/abs/hep-ph/9403398}{{\tt hep-ph/9403398}}].

\bibitem{Nardi:1994iv}
E.~Nardi, E.~Roulet, and D.~Tommasini, {\it {Limits on neutrino mixing with new
  heavy particles}},  {\em Phys. Lett. B} {\bf 327} (1994) 319--326,
  [\href{http://arxiv.org/abs/hep-ph/9402224}{{\tt hep-ph/9402224}}].

\bibitem{Tommasini:1995ii}
D.~Tommasini, G.~Barenboim, J.~Bernabeu, and C.~Jarlskog, {\it {Nondecoupling
  of heavy neutrinos and lepton flavor violation}},  {\em Nucl. Phys. B} {\bf
  444} (1995) 451--467, [\href{http://arxiv.org/abs/hep-ph/9503228}{{\tt
  hep-ph/9503228}}].

\bibitem{Illana:2000ic}
J.~I. Illana and T.~Riemann, {\it {Charged lepton flavor violation from massive
  neutrinos in Z decays}},  {\em Phys. Rev. D} {\bf 63} (2001) 053004,
  [\href{http://arxiv.org/abs/hep-ph/0010193}{{\tt hep-ph/0010193}}].

\bibitem{Loinaz:2003gc}
W.~Loinaz, N.~Okamura, S.~Rayyan, T.~Takeuchi, and L.~C.~R. Wijewardhana, {\it
  {Quark lepton unification and lepton flavor nonconservation from a TeV scale
  seesaw neutrino mass texture}},  {\em Phys. Rev. D} {\bf 68} (2003) 073001,
  [\href{http://arxiv.org/abs/hep-ph/0304004}{{\tt hep-ph/0304004}}].

\bibitem{Arganda:2004bz}
E.~Arganda, A.~M. Curiel, M.~J. Herrero, and D.~Temes, {\it {Lepton flavor
  violating Higgs boson decays from massive seesaw neutrinos}},  {\em Phys.
  Rev. D} {\bf 71} (2005) 035011,
  [\href{http://arxiv.org/abs/hep-ph/0407302}{{\tt hep-ph/0407302}}].

\bibitem{Loinaz:2004qc}
W.~Loinaz, N.~Okamura, S.~Rayyan, T.~Takeuchi, and L.~C.~R. Wijewardhana, {\it
  {The NuTeV anomaly, lepton universality, and nonuniversal neutrino gauge
  couplings}},  {\em Phys. Rev. D} {\bf 70} (2004) 113004,
  [\href{http://arxiv.org/abs/hep-ph/0403306}{{\tt hep-ph/0403306}}].

\bibitem{Antusch:2006vwa}
S.~Antusch, C.~Biggio, E.~Fernandez-Martinez, M.~B. Gavela, and J.~Lopez-Pavon,
  {\it {Unitarity of the Leptonic Mixing Matrix}},  {\em JHEP} {\bf 10} (2006)
  084, [\href{http://arxiv.org/abs/hep-ph/0607020}{{\tt hep-ph/0607020}}].

\bibitem{Antusch:2008tz}
S.~Antusch, J.~P. Baumann, and E.~Fernandez-Martinez, {\it {Non-Standard
  Neutrino Interactions with Matter from Physics Beyond the Standard Model}},
  {\em Nucl. Phys. B} {\bf 810} (2009) 369--388,
  [\href{http://arxiv.org/abs/0807.1003}{{\tt arXiv:0807.1003}}].

\bibitem{Biggio:2008in}
C.~Biggio, {\it {The Contribution of fermionic seesaws to the anomalous
  magnetic moment of leptons}},  {\em Phys. Lett. B} {\bf 668} (2008) 378--384,
  [\href{http://arxiv.org/abs/0806.2558}{{\tt arXiv:0806.2558}}].

\bibitem{Alonso:2012ji}
R.~Alonso, M.~Dhen, M.~B. Gavela, and T.~Hambye, {\it {Muon conversion to
  electron in nuclei in type-I seesaw models}},  {\em JHEP} {\bf 01} (2013)
  118, [\href{http://arxiv.org/abs/1209.2679}{{\tt arXiv:1209.2679}}].

\bibitem{Abada:2012mc}
A.~Abada, D.~Das, A.~M. Teixeira, A.~Vicente, and C.~Weiland, {\it {Tree-level
  lepton universality violation in the presence of sterile neutrinos: impact
  for $R_K$ and $R_\pi$}},  {\em JHEP} {\bf 02} (2013) 048,
  [\href{http://arxiv.org/abs/1211.3052}{{\tt arXiv:1211.3052}}].

\bibitem{Akhmedov:2013hec}
E.~Akhmedov, A.~Kartavtsev, M.~Lindner, L.~Michaels, and J.~Smirnov, {\it
  {Improving Electro-Weak Fits with TeV-scale Sterile Neutrinos}},  {\em JHEP}
  {\bf 05} (2013) 081, [\href{http://arxiv.org/abs/1302.1872}{{\tt
  arXiv:1302.1872}}].

\bibitem{Basso:2013jka}
L.~Basso, O.~Fischer, and J.~J. van~der Bij, {\it {Precision tests of unitarity
  in leptonic mixing}},  {\em EPL} {\bf 105} (2014), no.~1 11001,
  [\href{http://arxiv.org/abs/1310.2057}{{\tt arXiv:1310.2057}}].

\bibitem{Abada:2013aba}
A.~Abada, A.~M. Teixeira, A.~Vicente, and C.~Weiland, {\it {Sterile neutrinos
  in leptonic and semileptonic decays}},  {\em JHEP} {\bf 02} (2014) 091,
  [\href{http://arxiv.org/abs/1311.2830}{{\tt arXiv:1311.2830}}].

\bibitem{Arganda:2014dta}
E.~Arganda, M.~J. Herrero, X.~Marcano, and C.~Weiland, {\it {Imprints of
  massive inverse seesaw model neutrinos in lepton flavor violating Higgs boson
  decays}},  {\em Phys. Rev. D} {\bf 91} (2015), no.~1 015001,
  [\href{http://arxiv.org/abs/1405.4300}{{\tt arXiv:1405.4300}}].

\bibitem{Antusch:2014woa}
S.~Antusch and O.~Fischer, {\it {Non-unitarity of the leptonic mixing matrix:
  Present bounds and future sensitivities}},  {\em JHEP} {\bf 10} (2014) 094,
  [\href{http://arxiv.org/abs/1407.6607}{{\tt arXiv:1407.6607}}].

\bibitem{Antusch:2015mia}
S.~Antusch and O.~Fischer, {\it {Testing sterile neutrino extensions of the
  Standard Model at future lepton colliders}},  {\em JHEP} {\bf 05} (2015) 053,
  [\href{http://arxiv.org/abs/1502.05915}{{\tt arXiv:1502.05915}}].

\bibitem{Abada:2014cca}
A.~Abada, V.~De~Romeri, S.~Monteil, J.~Orloff, and A.~M. Teixeira, {\it
  {Indirect searches for sterile neutrinos at a high-luminosity Z-factory}},
  {\em JHEP} {\bf 04} (2015) 051, [\href{http://arxiv.org/abs/1412.6322}{{\tt
  arXiv:1412.6322}}].

\bibitem{Abada:2015oba}
A.~Abada, V.~De~Romeri, and A.~M. Teixeira, {\it {Impact of sterile neutrinos
  on nuclear-assisted cLFV processes}},  {\em JHEP} {\bf 02} (2016) 083,
  [\href{http://arxiv.org/abs/1510.06657}{{\tt arXiv:1510.06657}}].

\bibitem{Abada:2015trh}
A.~Abada and T.~Toma, {\it {Electric Dipole Moments of Charged Leptons with
  Sterile Fermions}},  {\em JHEP} {\bf 02} (2016) 174,
  [\href{http://arxiv.org/abs/1511.03265}{{\tt arXiv:1511.03265}}].

\bibitem{Fernandez-Martinez:2015hxa}
E.~Fernandez-Martinez, J.~Hernandez-Garcia, J.~Lopez-Pavon, and M.~Lucente,
  {\it {Loop level constraints on Seesaw neutrino mixing}},  {\em JHEP} {\bf
  10} (2015) 130, [\href{http://arxiv.org/abs/1508.03051}{{\tt
  arXiv:1508.03051}}].

\bibitem{Fernandez-Martinez:2016lgt}
E.~Fernandez-Martinez, J.~Hernandez-Garcia, and J.~Lopez-Pavon, {\it {Global
  constraints on heavy neutrino mixing}},  {\em JHEP} {\bf 08} (2016) 033,
  [\href{http://arxiv.org/abs/1605.08774}{{\tt arXiv:1605.08774}}].

\bibitem{Abada:2016awd}
A.~Abada and T.~Toma, {\it {Electron electric dipole moment in Inverse Seesaw
  models}},  {\em JHEP} {\bf 08} (2016) 079,
  [\href{http://arxiv.org/abs/1605.07643}{{\tt arXiv:1605.07643}}].

\bibitem{DeRomeri:2016gum}
V.~De~Romeri, M.~J. Herrero, X.~Marcano, and F.~Scarcella, {\it {Lepton flavor
  violating Z decays: A promising window to low scale seesaw neutrinos}},  {\em
  Phys. Rev. D} {\bf 95} (2017), no.~7 075028,
  [\href{http://arxiv.org/abs/1607.05257}{{\tt arXiv:1607.05257}}].

\bibitem{Arganda:2016zvc}
E.~Arganda, M.~J. Herrero, X.~Marcano, R.~Morales, and A.~Szynkman, {\it
  {Effective lepton flavor violating H\ensuremath{\ell}i\ensuremath{\ell}j
  vertex from right-handed neutrinos within the mass insertion approximation}},
   {\em Phys. Rev. D} {\bf 95} (2017), no.~9 095029,
  [\href{http://arxiv.org/abs/1612.09290}{{\tt arXiv:1612.09290}}].

\bibitem{Herrero:2018luu}
M.~J. Herrero, X.~Marcano, R.~Morales, and A.~Szynkman, {\it {One-loop
  effective LFV $Zl_kl_m$ vertex from heavy neutrinos within the mass insertion
  approximation}},  {\em Eur. Phys. J. C} {\bf 78} (2018), no.~10 815,
  [\href{http://arxiv.org/abs/1807.01698}{{\tt arXiv:1807.01698}}].

\bibitem{Marcano:2019rmk}
X.~Marcano and R.~A. Morales, {\it {Flavor techniques for LFV processes: Higgs
  decays in a general seesaw model}},  {\em Front. in Phys.} {\bf 7} (2020)
  228, [\href{http://arxiv.org/abs/1909.05888}{{\tt arXiv:1909.05888}}].

\bibitem{Chrzaszcz:2019inj}
M.~Chrzaszcz, M.~Drewes, T.~E. Gonzalo, J.~Harz, S.~Krishnamurthy, and
  C.~Weniger, {\it {A frequentist analysis of three right-handed neutrinos with
  GAMBIT}},  {\em Eur. Phys. J. C} {\bf 80} (2020), no.~6 569,
  [\href{http://arxiv.org/abs/1908.02302}{{\tt arXiv:1908.02302}}].

\bibitem{Coutinho:2019aiy}
A.~M. Coutinho, A.~Crivellin, and C.~A. Manzari, {\it {Global Fit to Modified
  Neutrino Couplings and the Cabibbo-Angle Anomaly}},  {\em Phys. Rev. Lett.}
  {\bf 125} (2020), no.~7 071802, [\href{http://arxiv.org/abs/1912.08823}{{\tt
  arXiv:1912.08823}}].

\bibitem{Calderon:2022alb}
K.~A. Urqu\'\i{}a-Calder\'on, I.~Timiryasov, and O.~Ruchayskiy, {\it {Heavy
  neutral leptons \textemdash{} Advancing into the PeV domain}},  {\em JHEP}
  {\bf 08} (2023) 167, [\href{http://arxiv.org/abs/2206.04540}{{\tt
  arXiv:2206.04540}}].

\bibitem{Blennow:2023mqx}
M.~Blennow, E.~Fern\'andez-Mart\'\i{}nez, J.~Hern\'andez-Garc\'\i{}a,
  J.~L\'opez-Pav\'on, X.~Marcano, and D.~Naredo-Tuero, {\it {Bounds on lepton
  non-unitarity and heavy neutrino mixing}},  {\em JHEP} {\bf 08} (2023) 030,
  [\href{http://arxiv.org/abs/2306.01040}{{\tt arXiv:2306.01040}}].

\bibitem{Bilenky:1992wv}
S.~M. Bilenky and C.~Giunti, {\it {Seesaw type mixing and muon-neutrino
  ---\ensuremath{>} tau-neutrino oscillations}},  {\em Phys. Lett. B} {\bf 300}
  (1993) 137--140, [\href{http://arxiv.org/abs/hep-ph/9211269}{{\tt
  hep-ph/9211269}}].

\bibitem{Bergmann:1998rg}
S.~Bergmann and A.~Kagan, {\it {Z - induced FCNCs and their effects on neutrino
  oscillations}},  {\em Nucl. Phys. B} {\bf 538} (1999) 368--386,
  [\href{http://arxiv.org/abs/hep-ph/9803305}{{\tt hep-ph/9803305}}].

\bibitem{Czakon:2001em}
M.~Czakon, J.~Gluza, and M.~Zralek, {\it {Nonunitary neutrino mixing matrix and
  CP violating neutrino oscillations}},  {\em Acta Phys. Polon. B} {\bf 32}
  (2001) 3735--3744, [\href{http://arxiv.org/abs/hep-ph/0109245}{{\tt
  hep-ph/0109245}}].

\bibitem{Bekman:2002zk}
B.~Bekman, J.~Gluza, J.~Holeczek, J.~Syska, and M.~Zralek, {\it {Matter effects
  and CP violating neutrino oscillations with nondecoupling heavy neutrinos}},
  {\em Phys. Rev. D} {\bf 66} (2002) 093004,
  [\href{http://arxiv.org/abs/hep-ph/0207015}{{\tt hep-ph/0207015}}].

\bibitem{Fernandez-Martinez:2007iaa}
E.~Fernandez-Martinez, M.~B. Gavela, J.~Lopez-Pavon, and O.~Yasuda, {\it
  {CP-violation from non-unitary leptonic mixing}},  {\em Phys. Lett. B} {\bf
  649} (2007) 427--435, [\href{http://arxiv.org/abs/hep-ph/0703098}{{\tt
  hep-ph/0703098}}].

\bibitem{Goswami:2008mi}
S.~Goswami and T.~Ota, {\it {Testing non-unitarity of neutrino mixing matrices
  at neutrino factories}},  {\em Phys. Rev. D} {\bf 78} (2008) 033012,
  [\href{http://arxiv.org/abs/0802.1434}{{\tt arXiv:0802.1434}}].

\bibitem{Antusch:2009pm}
S.~Antusch, M.~Blennow, E.~Fernandez-Martinez, and J.~Lopez-Pavon, {\it
  {Probing non-unitary mixing and CP-violation at a Neutrino Factory}},  {\em
  Phys. Rev. D} {\bf 80} (2009) 033002,
  [\href{http://arxiv.org/abs/0903.3986}{{\tt arXiv:0903.3986}}].

\bibitem{Meloni:2009cg}
D.~Meloni, T.~Ohlsson, W.~Winter, and H.~Zhang, {\it {Non-standard interactions
  versus non-unitary lepton flavor mixing at a neutrino factory}},  {\em JHEP}
  {\bf 04} (2010) 041, [\href{http://arxiv.org/abs/0912.2735}{{\tt
  arXiv:0912.2735}}].

\bibitem{Xing:2011ur}
Z.-z. Xing, {\it {A full parametrization of the 6 X 6 flavor mixing matrix in
  the presence of three light or heavy sterile neutrinos}},  {\em Phys. Rev. D}
  {\bf 85} (2012) 013008, [\href{http://arxiv.org/abs/1110.0083}{{\tt
  arXiv:1110.0083}}].

\bibitem{Forero:2011pc}
D.~V. Forero, S.~Morisi, M.~Tortola, and J.~W.~F. Valle, {\it {Lepton flavor
  violation and non-unitary lepton mixing in low-scale type-I seesaw}},  {\em
  JHEP} {\bf 09} (2011) 142, [\href{http://arxiv.org/abs/1107.6009}{{\tt
  arXiv:1107.6009}}].

\bibitem{Qian:2013ora}
X.~Qian, C.~Zhang, M.~Diwan, and P.~Vogel, {\it {Unitarity Tests of the
  Neutrino Mixing Matrix}},  \href{http://arxiv.org/abs/1308.5700}{{\tt
  arXiv:1308.5700}}.

\bibitem{Parke:2015goa}
S.~Parke and M.~Ross-Lonergan, {\it {Unitarity and the three flavor neutrino
  mixing matrix}},  {\em Phys. Rev. D} {\bf 93} (2016), no.~11 113009,
  [\href{http://arxiv.org/abs/1508.05095}{{\tt arXiv:1508.05095}}].

\bibitem{Escrihuela:2015wra}
F.~J. Escrihuela, D.~V. Forero, O.~G. Miranda, M.~Tortola, and J.~W.~F. Valle,
  {\it {On the description of nonunitary neutrino mixing}},  {\em Phys. Rev. D}
  {\bf 92} (2015), no.~5 053009, [\href{http://arxiv.org/abs/1503.08879}{{\tt
  arXiv:1503.08879}}]. [Erratum: Phys.Rev.D 93, 119905 (2016)].

\bibitem{Blennow:2016jkn}
M.~Blennow, P.~Coloma, E.~Fernandez-Martinez, J.~Hernandez-Garcia, and
  J.~Lopez-Pavon, {\it {Non-Unitarity, sterile neutrinos, and Non-Standard
  neutrino Interactions}},  {\em JHEP} {\bf 04} (2017) 153,
  [\href{http://arxiv.org/abs/1609.08637}{{\tt arXiv:1609.08637}}].

\bibitem{Ge:2016xya}
S.-F. Ge, P.~Pasquini, M.~Tortola, and J.~W.~F. Valle, {\it {Measuring the
  leptonic CP phase in neutrino oscillations with nonunitary mixing}},  {\em
  Phys. Rev. D} {\bf 95} (2017), no.~3 033005,
  [\href{http://arxiv.org/abs/1605.01670}{{\tt arXiv:1605.01670}}].

\bibitem{Escrihuela:2016ube}
F.~J. Escrihuela, D.~V. Forero, O.~G. Miranda, M.~T\'ortola, and J.~W.~F.
  Valle, {\it {Probing CP violation with non-unitary mixing in long-baseline
  neutrino oscillation experiments: DUNE as a case study}},  {\em New J. Phys.}
  {\bf 19} (2017), no.~9 093005, [\href{http://arxiv.org/abs/1612.07377}{{\tt
  arXiv:1612.07377}}].

\bibitem{Miranda:2016wdr}
O.~G. Miranda, M.~Tortola, and J.~W.~F. Valle, {\it {New ambiguity in probing
  CP violation in neutrino oscillations}},  {\em Phys. Rev. Lett.} {\bf 117}
  (2016), no.~6 061804, [\href{http://arxiv.org/abs/1604.05690}{{\tt
  arXiv:1604.05690}}].

\bibitem{Pas:2016qbg}
H.~P\"as and P.~Sicking, {\it {Discriminating sterile neutrinos and unitarity
  violation with CP invariants}},  {\em Phys. Rev. D} {\bf 95} (2017), no.~7
  075004, [\href{http://arxiv.org/abs/1611.08450}{{\tt arXiv:1611.08450}}].

\bibitem{Dutta:2016czj}
D.~Dutta, P.~Ghoshal, and S.~Roy, {\it {Effect of Non Unitarity on Neutrino
  Mass Hierarchy determination at DUNE, NO$\nu$A and T2K}},  {\em Nucl. Phys.
  B} {\bf 920} (2017) 385--401, [\href{http://arxiv.org/abs/1609.07094}{{\tt
  arXiv:1609.07094}}].

\bibitem{Bielas:2017lok}
K.~Bielas, W.~Flieger, J.~Gluza, and M.~Gluza, {\it {Neutrino mixing, interval
  matrices and singular values}},  {\em Phys. Rev. D} {\bf 98} (2018), no.~5
  053001, [\href{http://arxiv.org/abs/1708.09196}{{\tt arXiv:1708.09196}}].

\bibitem{Miranda:2018yym}
O.~G. Miranda, P.~Pasquini, M.~T\'ortola, and J.~W.~F. Valle, {\it {Exploring
  the Potential of Short-Baseline Physics at Fermilab}},  {\em Phys. Rev. D}
  {\bf 97} (2018), no.~9 095026, [\href{http://arxiv.org/abs/1802.02133}{{\tt
  arXiv:1802.02133}}].

\bibitem{Martinez-Soler:2018lcy}
I.~Martinez-Soler and H.~Minakata, {\it {Standard versus Non-Standard CP Phases
  in Neutrino Oscillation in Matter with Non-Unitarity}},  {\em PTEP} {\bf
  2020} (2020), no.~6 063B01, [\href{http://arxiv.org/abs/1806.10152}{{\tt
  arXiv:1806.10152}}].

\bibitem{Li:2018jgd}
Y.-F. Li, Z.-z. Xing, and J.-y. Zhu, {\it {Indirect unitarity violation
  entangled with matter effects in reactor antineutrino oscillations}},  {\em
  Phys. Lett. B} {\bf 782} (2018) 578--588,
  [\href{http://arxiv.org/abs/1802.04964}{{\tt arXiv:1802.04964}}].

\bibitem{Soumya:2018nkw}
C.~Soumya and M.~Rukmani, {\it {Non-unitary lepton mixing in an inverse seesaw
  and its impact on the physics potential of long-baseline experiments}},  {\em
  J. Phys. G} {\bf 45} (2018), no.~9 095003.

\bibitem{DeGouvea:2019kea}
A.~De~Gouv\^ea, K.~J. Kelly, G.~V. Stenico, and P.~Pasquini, {\it {Physics with
  Beam Tau-Neutrino Appearance at DUNE}},  {\em Phys. Rev. D} {\bf 100} (2019),
  no.~1 016004, [\href{http://arxiv.org/abs/1904.07265}{{\tt
  arXiv:1904.07265}}].

\bibitem{Escrihuela:2019mot}
F.~J. Escrihuela, L.~J. Flores, and O.~G. Miranda, {\it {Neutrino counting
  experiments and non-unitarity from LEP and future experiments}},  {\em Phys.
  Lett. B} {\bf 802} (2020) 135241,
  [\href{http://arxiv.org/abs/1907.12675}{{\tt arXiv:1907.12675}}].

\bibitem{Miranda:2019ynh}
L.~S. Miranda, P.~Pasquini, U.~Rahaman, and S.~Razzaque, {\it {Searching for
  non-unitary neutrino oscillations in the present T2K and NO$\nu $A data}},
  {\em Eur. Phys. J. C} {\bf 81} (2021), no.~5 444,
  [\href{http://arxiv.org/abs/1911.09398}{{\tt arXiv:1911.09398}}].

\bibitem{Ellis:2020ehi}
S.~A.~R. Ellis, K.~J. Kelly, and S.~W. Li, {\it {Leptonic Unitarity
  Triangles}},  {\em Phys. Rev. D} {\bf 102} (2020), no.~11 115027,
  [\href{http://arxiv.org/abs/2004.13719}{{\tt arXiv:2004.13719}}].

\bibitem{Ellis:2020hus}
S.~A.~R. Ellis, K.~J. Kelly, and S.~W. Li, {\it {Current and Future Neutrino
  Oscillation Constraints on Leptonic Unitarity}},  {\em JHEP} {\bf 12} (2020)
  068, [\href{http://arxiv.org/abs/2008.01088}{{\tt arXiv:2008.01088}}].

\bibitem{Miranda:2020syh}
O.~G. Miranda, D.~K. Papoulias, O.~Sanders, M.~T\'ortola, and J.~W.~F. Valle,
  {\it {Future CEvNS experiments as probes of lepton unitarity and
  light-sterile neutrinos}},  {\em Phys. Rev. D} {\bf 102} (2020) 113014,
  [\href{http://arxiv.org/abs/2008.02759}{{\tt arXiv:2008.02759}}].

\bibitem{Chakraborty:2020brc}
K.~Chakraborty, S.~Goswami, and K.~Long, {\it {New physics at nuSTORM}},  {\em
  Phys. Rev. D} {\bf 103} (2021), no.~7 075009,
  [\href{http://arxiv.org/abs/2007.03321}{{\tt arXiv:2007.03321}}].

\bibitem{Hu:2020oba}
Z.~Hu, J.~Ling, J.~Tang, and T.~Wang, {\it {Global oscillation data analysis on
  the $3\nu$ mixing without unitarity}},  {\em JHEP} {\bf 01} (2021) 124,
  [\href{http://arxiv.org/abs/2008.09730}{{\tt arXiv:2008.09730}}].

\bibitem{Coloma:2021uhq}
P.~Coloma, J.~L\'opez-Pav\'on, S.~Rosauro-Alcaraz, and S.~Urrea, {\it {New
  physics from oscillations at the DUNE near detector, and the role of
  systematic uncertainties}},  {\em JHEP} {\bf 08} (2021) 065,
  [\href{http://arxiv.org/abs/2105.11466}{{\tt arXiv:2105.11466}}].

\bibitem{Forero:2021azc}
D.~V. Forero, C.~Giunti, C.~A. Ternes, and M.~Tortola, {\it {Nonunitary
  neutrino mixing in short and long-baseline experiments}},  {\em Phys. Rev. D}
  {\bf 104} (2021), no.~7 075030, [\href{http://arxiv.org/abs/2103.01998}{{\tt
  arXiv:2103.01998}}].

\bibitem{Denton:2021mso}
P.~B. Denton and J.~Gehrlein, {\it {New oscillation and scattering constraints
  on the tau row matrix elements without assuming unitarity}},  {\em JHEP} {\bf
  06} (2022) 135, [\href{http://arxiv.org/abs/2109.14575}{{\tt
  arXiv:2109.14575}}].

\bibitem{Denton:2021rsa}
P.~B. Denton, {\it {Tau neutrino identification in atmospheric neutrino
  oscillations without particle identification or unitarity}},  {\em Phys. Rev.
  D} {\bf 104} (2021), no.~11 113003,
  [\href{http://arxiv.org/abs/2109.14576}{{\tt arXiv:2109.14576}}].

\bibitem{Agarwalla:2021owd}
S.~K. Agarwalla, S.~Das, A.~Giarnetti, and D.~Meloni, {\it {Model-independent
  constraints on non-unitary neutrino mixing from high-precision long-baseline
  experiments}},  {\em JHEP} {\bf 07} (2022) 121,
  [\href{http://arxiv.org/abs/2111.00329}{{\tt arXiv:2111.00329}}].

\bibitem{Denton:2022pxt}
P.~B. Denton, A.~Giarnetti, and D.~Meloni, {\it {How to identify different new
  neutrino oscillation physics scenarios at DUNE}},  {\em JHEP} {\bf 02} (2023)
  210, [\href{http://arxiv.org/abs/2210.00109}{{\tt arXiv:2210.00109}}].

\bibitem{Fong:2022oim}
C.~S. Fong, {\it {Analytic neutrino oscillation probabilities}},  {\em SciPost
  Phys.} {\bf 15} (2023), no.~1 013,
  [\href{http://arxiv.org/abs/2210.09436}{{\tt arXiv:2210.09436}}].

\bibitem{Aloni:2022ebm}
D.~Aloni and A.~Dery, {\it {Revisiting leptonic nonunitarity}},  {\em Phys.
  Rev. D} {\bf 109} (2024), no.~5 055006,
  [\href{http://arxiv.org/abs/2211.09638}{{\tt arXiv:2211.09638}}].

\bibitem{Fong:2023fpt}
C.~S. Fong, {\it {Theoretical Aspect of Nonunitarity in Neutrino Oscillation}},
   \href{http://arxiv.org/abs/2301.12960}{{\tt arXiv:2301.12960}}.

\bibitem{Fong:2023ams}
C.~S. Fong, {\it {New Physics in Neutrino Oscillation: Nonunitarity or
  Nonorthogonality?}},  \href{http://arxiv.org/abs/2305.19755}{{\tt
  arXiv:2305.19755}}.

\bibitem{Celestino-Ramirez:2023zox}
J.~M. Celestino-Ram\'\i{}rez, F.~J. Escrihuela, L.~J. Flores, and O.~G.
  Miranda, {\it {Testing the nonunitarity of the leptonic mixing matrix at
  FASERv and FASERv2}},  {\em Phys. Rev. D} {\bf 109} (2024), no.~1 L011705,
  [\href{http://arxiv.org/abs/2309.00116}{{\tt arXiv:2309.00116}}].

\bibitem{Sahoo:2023mpj}
S.~Sahoo, S.~Das, A.~Kumar, and S.~K. Agarwalla, {\it {Constraining non-unitary
  neutrino mixing using matter effects in atmospheric neutrinos at INO-ICAL}},
  {\em JHEP} {\bf 09} (2024) 184, [\href{http://arxiv.org/abs/2309.16942}{{\tt
  arXiv:2309.16942}}].

\bibitem{Kozynets:2024xgt}
T.~Kozynets, P.~Eller, A.~Zander, M.~Ettengruber, and D.~J. Koskinen, {\it
  {Constraints on non-unitary neutrino mixing in light of atmospheric and
  reactor neutrino data}},  \href{http://arxiv.org/abs/2407.20388}{{\tt
  arXiv:2407.20388}}.

\bibitem{Trzeciak:2025hap}
A.~M.~G. Trzeciak, H.~Nunokawa, and A.~A. Quiroga, {\it {Impact of unitarity
  violation on sensitivity of the leptonic CP phase at Hyper-Kamiokande and
  DUNE}},  \href{http://arxiv.org/abs/2502.10873}{{\tt arXiv:2502.10873}}.

\bibitem{Fong:2016yyh}
C.~S. Fong, H.~Minakata, and H.~Nunokawa, {\it {A framework for testing
  leptonic unitarity by neutrino oscillation experiments}},  {\em JHEP} {\bf
  02} (2017) 114, [\href{http://arxiv.org/abs/1609.08623}{{\tt
  arXiv:1609.08623}}].

\bibitem{Xing:2007zj}
Z.-z. Xing, {\it {Correlation between the Charged Current Interactions of Light
  and Heavy Majorana Neutrinos}},  {\em Phys. Lett. B} {\bf 660} (2008)
  515--521, [\href{http://arxiv.org/abs/0709.2220}{{\tt arXiv:0709.2220}}].

\bibitem{Coloma:2024ict}
P.~Coloma, E.~Fern\'andez-Mart\'\i{}nez, J.~L\'opez-Pav\'on, X.~Marcano,
  D.~Naredo-Tuero, and S.~Urrea, {\it {Improving the Global SMEFT Picture with
  Bounds on Neutrino NSI}},  {\em JHEP} {\bf 02} (2025) 137,
  [\href{http://arxiv.org/abs/2411.00090}{{\tt arXiv:2411.00090}}].

\bibitem{Eitel:2000by}
{\bf KARMEN} Collaboration, K.~Eitel, {\it {Latest results of the KARMEN2
  experiment}},  {\em Nucl. Phys. B Proc. Suppl.} {\bf 91} (2001) 191--197,
  [\href{http://arxiv.org/abs/hep-ex/0008002}{{\tt hep-ex/0008002}}].

\bibitem{NOMAD:2001xxt}
{\bf NOMAD} Collaboration, P.~Astier et~al., {\it {Final NOMAD results on
  muon-neutrino ---\ensuremath{>} tau-neutrino and electron-neutrino
  ---\ensuremath{>} tau-neutrino oscillations including a new search for
  tau-neutrino appearance using hadronic tau decays}},  {\em Nucl. Phys. B}
  {\bf 611} (2001) 3--39, [\href{http://arxiv.org/abs/hep-ex/0106102}{{\tt
  hep-ex/0106102}}].

\bibitem{NOMAD:2003mqg}
{\bf NOMAD} Collaboration, P.~Astier et~al., {\it {Search for nu(mu)
  ---\ensuremath{>} nu(e) oscillations in the NOMAD experiment}},  {\em Phys.
  Lett. B} {\bf 570} (2003) 19--31,
  [\href{http://arxiv.org/abs/hep-ex/0306037}{{\tt hep-ex/0306037}}].

\bibitem{Donini:2011jh}
A.~Donini, P.~Hernandez, J.~Lopez-Pavon, and M.~Maltoni, {\it {Minimal models
  with light sterile neutrinos}},  {\em JHEP} {\bf 07} (2011) 105,
  [\href{http://arxiv.org/abs/1106.0064}{{\tt arXiv:1106.0064}}].

\bibitem{Donini:2012tt}
A.~Donini, P.~Hernandez, J.~Lopez-Pavon, M.~Maltoni, and T.~Schwetz, {\it {The
  minimal 3+2 neutrino model versus oscillation anomalies}},  {\em JHEP} {\bf
  07} (2012) 161, [\href{http://arxiv.org/abs/1205.5230}{{\tt
  arXiv:1205.5230}}].

\bibitem{Goldhagen:2021kxe}
K.~Goldhagen, M.~Maltoni, S.~E. Reichard, and T.~Schwetz, {\it {Testing sterile
  neutrino mixing with present and future solar neutrino data}},  {\em Eur.
  Phys. J. C} {\bf 82} (2022), no.~2 116,
  [\href{http://arxiv.org/abs/2109.14898}{{\tt arXiv:2109.14898}}].

\bibitem{Stuttard:2020zsj}
{\bf IceCube} Collaboration, T.~Stuttard, {\it {Neutrino oscillations and PMNS
  unitarity with IceCube/DeepCore and the IceCube Upgrade}},  {\em PoS} {\bf
  NuFact2019} (2020) 099.

\bibitem{IceCube:2024dlz}
{\bf IceCube} Collaboration, R.~Abbasi et~al., {\it {Search for a light sterile
  neutrino with 7.5~years of IceCube DeepCore data}},  {\em Phys. Rev. D} {\bf
  110} (2024), no.~7 072007, [\href{http://arxiv.org/abs/2407.01314}{{\tt
  arXiv:2407.01314}}].

\bibitem{KM3NeT:2025ftj}
{\bf KM3NeT} Collaboration, S.~Aiello et~al., {\it {Study of tau neutrinos and
  non-unitary neutrino mixing with the first six detection units of
  KM3NeT/ORCA}},  \href{http://arxiv.org/abs/2502.01443}{{\tt
  arXiv:2502.01443}}.

\bibitem{Blennow:2018hto}
M.~Blennow, E.~Fernandez-Martinez, J.~Gehrlein, J.~Hernandez-Garcia, and
  J.~Salvado, {\it {IceCube bounds on sterile neutrinos above 10 eV}},  {\em
  Eur. Phys. J. C} {\bf 78} (2018), no.~10 807,
  [\href{http://arxiv.org/abs/1803.02362}{{\tt arXiv:1803.02362}}].

\bibitem{Dziewonski:1981xy}
A.~M. Dziewonski and D.~L. Anderson, {\it {Preliminary reference earth model}},
   {\em Phys. Earth Planet. Interiors} {\bf 25} (1981) 297--356.

\bibitem{Coloma:2022umy}
P.~Coloma, M.~C. Gonzalez-Garcia, M.~Maltoni, J.~a.~P. Pinheiro, and S.~Urrea,
  {\it {Constraining new physics with Borexino Phase-II spectral data}},  {\em
  JHEP} {\bf 07} (2022) 138, [\href{http://arxiv.org/abs/2204.03011}{{\tt
  arXiv:2204.03011}}]. [Erratum: JHEP 11, 138 (2022)].

\bibitem{Dutta:2019hmb}
D.~Dutta and S.~Roy, {\it {Non-Unitarity at DUNE and T2HK with Charged and
  Neutral Current Measurements}},  {\em J. Phys. G} {\bf 48} (2021), no.~4
  045004, [\href{http://arxiv.org/abs/1901.11298}{{\tt arXiv:1901.11298}}].

\bibitem{Acero:2022wqg}
M.~A. Acero et~al., {\it {White paper on light sterile neutrino searches and
  related phenomenology}},  {\em J. Phys. G} {\bf 51} (2024), no.~12 120501,
  [\href{http://arxiv.org/abs/2203.07323}{{\tt arXiv:2203.07323}}].

\bibitem{MINOS:2020iqj}
{\bf MINOS+, Daya Bay} Collaboration, P.~Adamson et~al., {\it {Improved
  Constraints on Sterile Neutrino Mixing from Disappearance Searches in the
  MINOS, MINOS+, Daya Bay, and Bugey-3 Experiments}},  {\em Phys. Rev. Lett.}
  {\bf 125} (2020), no.~7 071801, [\href{http://arxiv.org/abs/2002.00301}{{\tt
  arXiv:2002.00301}}].

\bibitem{Dentler:2018sju}
M.~Dentler, A.~Hern\'andez-Cabezudo, J.~Kopp, P.~A.~N. Machado, M.~Maltoni,
  I.~Martinez-Soler, and T.~Schwetz, {\it {Updated Global Analysis of Neutrino
  Oscillations in the Presence of eV-Scale Sterile Neutrinos}},  {\em JHEP}
  {\bf 08} (2018) 010, [\href{http://arxiv.org/abs/1803.10661}{{\tt
  arXiv:1803.10661}}].

\bibitem{Atre:2009rg}
A.~Atre, T.~Han, S.~Pascoli, and B.~Zhang, {\it {The Search for Heavy Majorana
  Neutrinos}},  {\em JHEP} {\bf 05} (2009) 030,
  [\href{http://arxiv.org/abs/0901.3589}{{\tt arXiv:0901.3589}}].

\bibitem{Fernandez-Martinez:2023phj}
E.~Fern\'andez-Mart\'\i{}nez, M.~Gonz\'alez-L\'opez,
  J.~Hern\'andez-Garc\'\i{}a, M.~Hostert, and J.~L\'opez-Pav\'on, {\it
  {Effective portals to heavy neutral leptons}},  {\em JHEP} {\bf 09} (2023)
  001, [\href{http://arxiv.org/abs/2304.06772}{{\tt arXiv:2304.06772}}].

\end{thebibliography}\endgroup

\end{document}